\documentclass[fleqn,usenatbib]{mnras}

\usepackage{graphicx}	
\usepackage{amsmath,bm}	
\usepackage[dvipsnames]{xcolor}
\usepackage{hyperref}
\usepackage{newtxtext,newtxmath}
\usepackage[T1]{fontenc}
\usepackage{adjustbox}
\usepackage{upgreek}
\usepackage{ulem,cancel}
\usepackage{nicefrac, xfrac}
\usepackage{academicons}
\DeclareRobustCommand{\VAN}[3]{#2}
\let\VANthebibliography\thebibliography
\def\thebibliography{\DeclareRobustCommand{\VAN}[3]{##3}\VANthebibliography}

\definecolor{midblue}{rgb}{0.0,0.4,0.7}
\definecolor{mypurple}{rgb}{0.7,0.3,0.8}

\definecolor{PineGreen}{HTML}{008B72}
\definecolor{Berry}{HTML}{FF2052}

\newcommand{\gafd}{Geophys.\ Astrophys.\ Fluid Dyn.}

\newcommand{\RThA}{{\tt{R3h2}}}
\newcommand{\RThB}{{\tt{R3h3}}}
\newcommand{\RThC}{{\tt{R3h6}}}

\newcommand{\RVhA}{{\tt{R5h2}}}
\newcommand{\RVhB}{{\tt{R5h3}}}
\newcommand{\RVhC}{{\tt{R5h6}}}

\newcommand{\RHhA}{{\tt{R7h2}}}
\newcommand{\RHhB}{{\tt{R7h3}}}
\newcommand{\RHhC}{{\tt{R7h6}}}
\newcommand{\RXhA}{{\tt{R10h2}}}
\newcommand{\RXhB}{{\tt{R10h3}}}
\newcommand{\RXhC}{{\tt{R10h6}}}
\newcommand{\RXVhA}{{\tt{R15h2}}}
\newcommand{\RXVhB}{{\tt{R15h3}}}
\newcommand{\RXVhC}{{\tt{R15h6}}}
\newcommand{\RXXhA}{{\tt{R20h2}}}
\newcommand{\RXXhB}{{\tt{R20h3}}}
\newcommand{\RXXhC}{{\tt{R20h6}}}

\newcommand{\RThAb}{{\tt{R3h2b}}}
\newcommand{\RThBb}{{\tt{R3h3b}}}
\newcommand{\RThCb}{{\tt{R3h6b}}}

\newcommand{\RVhAb}{{\tt{R5h2b}}}
\newcommand{\RVhBb}{{\tt{R5h3b}}}
\newcommand{\RVhCb}{{\tt{R5h6b}}}

\newcommand{\RHhAb}{{\tt{R7h2b}}}
\newcommand{\RHhBb}{{\tt{R7h3b}}}
\newcommand{\RHhCb}{{\tt{R7h6b}}}
\newcommand{\RXhAb}{{\tt{R10h2b}}}
\newcommand{\RXhBb}{{\tt{R10h3b}}}
\newcommand{\RXhCb}{{\tt{R10h6b}}}
\newcommand{\RXVhAb}{{\tt{R15h2b}}}
\newcommand{\RXVhBb}{{\tt{R15h3b}}}
\newcommand{\RXVhCb}{{\tt{R15h6b}}}
\newcommand{\RXXhAb}{{\tt{R20h2b}}}
\newcommand{\RXXhBb}{{\tt{R20h3b}}}
\newcommand{\RXXhCb}{{\tt{R20h6b}}}

\newcommand{\dd}{\mathrm{d}}        
\newcommand\sound{_\text{s}} 
\newcommand\deriv[2]{\frac{\partial#1}{\partial#2}}
\newcommand\dderiv[2]{\frac{\text{D}#1}{\text{D}#2}}
\newcommand\mean[1]{\langle #1\rangle}
\newcommand\meanh[1]{{\langle #1\rangle}_{xy}}
\renewcommand\vec[1]{\bm{#1}}
\newcommand\kin{_\text{k}} 
\newcommand\magn{_\text{m}} 
\newcommand{\BB}{\vec{B}} 
\newcommand{\U}{\vec{u}} 
\renewcommand{\mathbfss}[1]{{{\mbox{\boldmath{$#1$}}}}}

\DeclareMathAlphabet{\mathsc}{OT1}{cmr}{m}{sc}
\def\testbx{bx}%
\DeclareRobustCommand{\ion}[2]{%
\relax\ifmmode
\ifx\testbx\f@series
{\mathbf{#1\,\mathsc{#2}}}\else
{\mathrm{#1\,\mathsc{#2}}}\fi
\else\textup{#1\,{\mdseries\textsc{#2}}}%
\fi}

\newcommand{\cm}{\,{\rm cm}}    
\newcommand{\km}{\,{\rm km}}    
\newcommand{\p}{\,{\rm pc}}     
\newcommand{\kpc}{\,{\rm kpc}}  
\newcommand{\s}{\,{\rm s}}      
\newcommand{\yr}{\,{\rm yr}}    
\newcommand{\Myr}{\,{\rm Myr}} 
\newcommand{\Gyr}{\,{\rm Gyr}}  
\newcommand{\kms}{\km\s^{-1}}    
\newcommand{\mkG}{\,\upmu{\rm G}} 
\newcommand{\K}{\,{\rm K}}      

\title[Nonlinear magnetic buoyancy and dynamo]{Nonlinear magnetic buoyancy instability
and turbulent dynamo}

\author[Y.~Qazi et al.]{Yasin Qazi,$^{1}$\thanks{E-mail: Y.Qazi@newcastle.ac.uk}\href{https://orcid.org/0009-0008-9513-8761}{\includegraphics[scale=0.5]{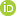}}
Anvar Shukurov,$^{1}$\href{https://orcid.org/0000-0001-6200-4304}{\includegraphics[scale=0.5]{orcid_16x16.jpeg}}
Devika Tharakkal,$^{1,2}$\href{https://orcid.org/0000-0002-4563-2277}{\includegraphics[scale=0.5]{orcid_16x16.jpeg}}
\newauthor
Frederick A.~Gent$^{3,4,1}$\href{https://orcid.org/0000-0002-1331-2260}{\includegraphics[scale=0.5]{orcid_16x16.jpeg}}
\&
Abhijit B.~Bendre$^{5}$\href{https://orcid.org/0000-0001-5208-8989}{\includegraphics[scale=0.5]{orcid_16x16.jpeg}}
\\
$^{1}$School of Mathematics, Statistics and Physics, Newcastle University, Newcastle upon Tyne, NE1 7RU, UK\\
$^{2}$Department of Physics, University of Helsinki, PO Box 64, FI-00014, Helsinki, Finland\\
$^{3}$Astroinformatics, Department of Computer Science, Aalto University, PO Box 15400, FI-00076, Espoo, Finland\\
$^{4}$Nordita, KTH Royal Institute of Technology and Stockholm University, Hannes Alfv\'ens v\"ag 12, Stockholm, SE-106, Sweden\\
$^{5}$Laboratoire d’Astrophysique, EPFL, CH-1290 Sauverny, Switzerland
}
\date{Accepted XXX. Received YYY; in original form ZZZ}

\pubyear{2023}

\begin{document}
\label{firstpage}
\pagerange{\pageref{firstpage}--\pageref{lastpage}}
\maketitle

\begin{abstract}
Stratified disks with strong horizontal magnetic fields,
are susceptible to magnetic buoyancy instability
(MBI). Modifying the magnetic field and gas distributions, this can play an
important role in galactic evolution. The MBI and the Parker instability, in
which MBI is exacerbated by cosmic rays, are often studied using an imposed
magnetic field. However, in galaxies and accretion discs, the magnetic field is
continuously replenished by a large-scale dynamo action. Using
non-ideal MHD equations, we model a section of the galactic disc
(we neglect rotation and cosmic rays considered elsewhere), 
in which the large-scale field is generated by an
imposed $\alpha$-effect of variable intensity to explore the interplay between
dynamo instability and MBI. The system evolves through three distinct phases:
the linear (kinematic) dynamo stage, the onset of linear MBI when the magnetic
field becomes sufficiently strong and the nonlinear, statistically steady state.
Nonlinear effects associated with the MBI introduce oscillations which do not occur
when the field is produced by the dynamo alone.
The MBI initially accelerates the magnetic field
amplification but the growth is quenched by the
vertical motions produced by MBI.
We construct a 1D model, which
replicates all significant features of 3D simulations
to confirm that magnetic buoyancy alone can quench the dynamo and is
responsible for the magnetic field oscillations.
Unlike similar results obtained with an imposed magnetic field,
the nonlinear interactions do not reduce the gas scale height,
so the consequences of the magnetic buoyancy depend on how the magnetic field is maintained.
\end{abstract}


\section{Introduction}\label{sec:intro}


Magnetic fields are present in many astrophysical systems, often exhibiting
highly random structure driven by turbulent flows.  It is also very common to
find magnetic fields coherently organised on large scales, partially aligned to
large-scale shear flows in rotating objects such as spiral galaxies, convecting
stars or accretion disks.  The large-scale magnetic fields are produced and
maintained by dynamo processes which amplify weak seed magnetic fields
\citep[Chapter~9 of][]{SS21} to generate dynamically significant large-scale
magnetic fields \citep{Moffatt1978, Parker1979, 1980mfmd_Krause_Radler}.

The linear (kinematic) stage of the magnetic field amplification, when the
Lorentz force is negligible and the magnetic fields still do not affect plasma
flows, is well understood but the nonlinear behaviours of magnetic fields and
plasma flows remain a subject of active research \citep{Brandenburg_2005,
SS21}.  Magnetic fields produced by the mean-field dynamo are subject to
further instabilities, and the magnetic buoyancy instability (MBI) is one of
the strongest and most ubiquitous of them. The dynamo and magnetic buoyancy
instabilities are most often studied separately although it is understood that
they can affect each other. In particular, plasma flows produced by the MBI
become helical under the action of the Coriolis force, which leads to the
mean-field dynamo action \citep{1992PARKER, 1998Han&Les, 1999MOSS,
Thelen_aw_2000, DT2022b} in addition to the dynamo associated  with the
turbulence driven in galaxies by supernova explosions. As we discuss below, the
dynamo action also affects the MBI. It is difficult to distinguish the effects
of the mean-field dynamo and magnetic buoyancy as they become strongly
interdependent in the nonlinear state.

The MBI is expected to develop in any magnetized, strongly stratified
astrophysical object and has a relatively short time scale of the order of the
sound or Alfv\'en crossing time based on the magnetic field scale height, of
order $10^7\yr$ in the Solar neighbourhood of the Milky Way \citep[Section
2.8.2 of][and references therein]{SS21}.  In spiral galaxies, the MBI enhanced
by cosmic rays (which produce significant pressure but have negligible weight)
is known as the Parker instability \citep{1966ApJ_145_811P}.

The linear stage of the MBI has been extensively explored with and without
cosmic rays but its nonlinear states have attracted less attention. The Parker
instability has been studied by \citet{1998Han&Les}, \citet{HO-ML02, HKO-ML04}
and \citet{KH07} with emphasis on the associated mean-field dynamo action,
including the nonlinear steady state. In application to accretion discs, the
role of the MBI and its effect on the magnetic field structure is discussed by
\citet{JoLe08} and \citet{GJL12}.

More recently, the nonlinear magnetic buoyancy and Parker instabilities of an
imposed azimuthal magnetic field has been examined by \citet{DT2022a, DT2022b}.
They show that magnetic buoyancy in a non-rotating system leads to a
substantial redistribution of gas, magnetic field and cosmic rays, resulting in
large (a few kiloparsecs) scale heights of the magnetic field and cosmic rays
and a correspondingly small (of order a hundred parsec) scale height of the
gas, which is supported against gravity mainly by thermal pressure.  Rotation
leads to somewhat smaller scale heights of the nonthermal constituents (and
their larger contribution to the pressure gradient near the midplane). More
importantly, the nonlinear MBI in a rotating system can completely change the
initial magnetic field even leading to its reversal.

A natural question to ask then is whether a hydromagnetic dynamo can sustain
the magnetic field against these effects and how this affects the evolution of
the instability. How might this change our understanding of galactic dynamos
and the structure of galactic magnetic fields?

To address such questions, we study the nonlinear states of the mean-field
dynamo in a disc, paying attention to the effects of the MBI. We explore a
range of models with various relations between the dynamo and MBI time scales.
The mean-field dynamo action is introduced via an explicit $\alpha$-effect in
the induction equation, assumed to be driven by background turbulent flows.
Anticipating that rotation would introduce significant qualitative complexity
(particularly the enhanced dynamo action arising from the motions induced by
the magnetic buoyancy), we commence with a simpler model in which we neglect
the overall rotation.  Although the $\alpha$-effect emerges physically only in
a rotating system, the specific effects of rotation is our subject for a
subsequent paper.

\section{Basic equations and numerical methods}\label{sec:methods}

We model isothermal gas and magnetic field in a section of a 
within a Cartesian box, with $x$, $y$ and $z$ representing the radial,
azimuthal and vertical directions, respectively. The simulation domain extends
$4\kpc$ in each horizontal direction and $3\kpc$ vertically, centered at the
galactic midplane.  We have tested computational boxes of various sizes from
$0.5$ to $8\kpc$ to confirm that we capture all essential features of the
evolving system.  The grid resolution is $256\times256\times384$ mesh points,
yielding the grid spacing about $15.6\p$ along each dimension. The size of the
domain is larger than the expected vertical and horizontal scales of the
instability, and the resolution is sufficient to obtain convergent solutions to
model the instability \citep{DT2022a}.

\begin{table}
\caption{
Parameters adopted in the numerical solutions of
equations~\eqref{eq:mass_conservation}--\eqref{eq:induction}.
\label{tab:parameter_values}
}
\centering
\begin{tabular}{@{}llll@{}}
\hline
Quantity                   & Symbol          & Value                                 &Unit     \\
\hline
Grid resolution            & $\updelta\vec{x}$    & 0.0156                                &kpc \\
Sound speed                & $c\sound$              & 15                                    &km$\s^{-1}$ \\
Kinematic viscosity        & $\nu$              & 0.008                         &kpc km$\s^{-1}$ \\
Magnetic diffusivity       & $\eta$             & 0.03                          &kpc km$\s^{-1}$ \\
$\alpha$-effect magnitude          & $\alpha_0$         & 0.15--3.0             &km$\s^{-1}$  \\
Dynamo scale height        & $h_\alpha$         & 0.2--0.6                      &kpc\\
Magnetic diffusion time     & $h_\alpha^2/\eta$  & 4/3--12                      &Gyr\\
Initial gas column density     &$\Sigma$   & $10^{21}$   & cm$^{-2}$\\
Shock-capturing viscosity   & $\nu_\text{shock}$& 12    & kpc$^{2}$\\
Shock-capturing diffusivity & $D_\text{shock}$  &  12   & kpc$^{2}$\\
Hyper-diffusivity            & $\nu_6,\ \eta_6$     & $10^{-10}$                &kpc$^{5}\kms$\\
\hline
\end{tabular}
\end{table}
\label{sec:model_desc}
\subsection{Basic equations}\label{sec:eqs}

We solve the system of isothermal non-ideal compressible MHD equations using
the sixth-order in space and third-order in time finite-difference
\textsc{Pencil Code} \citep{brandenburg2002,Pencil-JOSS},
\begin{align}
    \dderiv{\rho}{t} &= -\rho\nabla \cdot \U +\nabla \cdot\zeta_D\nabla\rho\,,
            \label{eq:mass_conservation}\\
    \dderiv{\U}{t} &= g\hat{\vec{z}} - c\sound^2\frac{\nabla\rho}{\rho}
        +\frac{(\nabla\times\BB)\times\BB}{4\pi\rho}
        +\frac{1}{\rho} \nabla\cdot(2\rho\nu{\mathbfss{\uptau}})
       \nonumber\\
    \label{eq:momentum}
        & +\nabla\left(\zeta_{\nu}\nabla \cdot \U \right)
        +\nabla\cdot \left(2\rho\nu_6{\mathbfss{\uptau}}^{(5)}\right)
        -\dfrac{1}{\rho}\U\nabla\cdot\left(\zeta_D\nabla\rho\right)\,,\\
    \deriv{\vec{A}}{t} &=  \alpha \BB+ \U\times \BB -\eta \nabla\times \BB
    \label{eq:induction}
    +\eta_6\nabla^{(6)}\vec{A}\,,
\end{align}
where $\text{D}/\text{D}t=\partial/\partial t + \U\cdot\nabla$, $\rho$ and $\U$
are the gas density and velocity, $\vec{A}$ is the magnetic vector potential,
with $\BB=\nabla \times \vec{A}$ for the magnetic field (thus, $\nabla \cdot
\BB \equiv 0$), and $c\sound$ is the speed of sound assumed to be constant.
equation~\eqref{eq:induction} includes the term $\alpha\vec{B}$ responsible for
the mean-field dynamo action, where $\alpha$ is specified in
Section~\ref{sec:dom_and_init}.  The key parameters used are shown in
Table~\ref{tab:parameter_values}.

The traceless rate of strain tensor $\mathbfss{\uptau}$ has the form
$\uptau_{ij}=\tfrac{1}{2}\,(\partial_j{u_i}+\partial_i{u_j})
-\tfrac{1}{3}\,\delta_{ij}\partial_k{u_k}$ (where $\partial_i=\partial/\partial
x_i$ and summation over repeated indices is understood). The shock viscosity
$\zeta_\nu = \nu_\text{shock}f_\text{shock}$ in equation~\eqref{eq:momentum},
with $f_\text{shock} \propto |\nabla\cdot\U_{-}|$ (where  $\nu_\text{shock}$ is
a constant given in Table~\ref{sec:model_desc} and $\nabla\cdot\U_{-}$ is the
divergence of the velocity field where it is negative and vanishes otherwise)
regularises shock fronts propagating perpendicular to steep pressure gradients,
and differs from zero only in regions of converging flow \citep[see][for
details]{GMKSH20}.  Following \citet{GMKSH20}, we also include the terms with
$\zeta_D$ in equations~\eqref{eq:mass_conservation} and \eqref{eq:momentum} to
ensure the momentum conservation.  Similarly to $\zeta_{\nu}$, the
shock-capturing term in the continuity equation has $\zeta_D =
D_\text{shock}f_\text{shock}$, where $D_\text{shock}$ is a constant given in
Table~\ref{sec:model_desc}.  The hyper diffusion with the coefficients $\nu_6$
and $\eta_6$ is used to resolve grid-scale numerical instabilities, with
$\uptau_{ij}^{(5)}=\tfrac{1}{2}\left[\partial_i^5 u_j
+\partial_i^{4}(\partial_j{u_i})\right]
-\tfrac{1}{6}\partial_i^4(\delta_{ij}\partial_k{u_k})$ and
$\nabla^{(6)}A_i=\partial_j^3\partial_j^3 A_i$, where $\partial_i^n =
\partial^n/\partial x_i^n$ \citep{ABGS02, GMKS21}.

\subsection{Boundary conditions}\label{sec:bound_cond}

The boundary conditions in both horizontal directions $x$ and $y$ are periodic
for all variables.  To prevent an artificial inward advection of the magnetic
energy through the top and bottom of the domain at $z=\pm1.5\kpc$, we impose
the conditions $B_x=B_y=0$ and $\partial B_z/\partial z=0$, i.e.,
\begin{equation}
   \partial A_x/\partial z = \partial A_y/\partial z = A_z = 0 \quad \text{at\ } |z| = 1.5\kpc\,.
\end{equation}
The boundary conditions for the velocity field are stress-free,
\begin{equation}
    \partial u_x/\partial z= \partial u_y/\partial z= 0 \quad \text{at\ } |z| = 1.5\kpc\,.
\end{equation}
To permit vertical gas flow across the boundaries without exciting numerical
instabilities, the boundary condition for $u_z$ imposes the boundary outflow
speed across the ghost zones outside the domain whereas an inflow speed  at the
boundary  tends smoothly to zero across the ghost zones \citep{Gent_SN_ISM_1}.
The density gradient is kept at a constant level at the boundaries, with the
scale height intermediate between that of the Lockman layer and the galactic
halo,
\begin{equation}
    \frac{\partial \ln\rho}{\partial z} = \pm\frac{1}{0.9\kpc} \quad \text{at\ }  z = \mp 1.5\kpc\,,
\end{equation}
and we note that the value of the scale height adopted has a negligible effect
on the results.

\subsection{The stratified gas layer}\label{sec:dom_and_init}

The gas is stratified by the application of  vertical gravitational
acceleration due to the stellar disc and dark matter halo following
\citet{K&G1989MNRAS} \citep[see also][]{Ferriere_1998},
\begin{equation}
    g = -a_1\frac{ z}{\sqrt{{z_1}^2+z^2}} -a_2\frac{z}{z_2}\,,
    \label{eq:acceleration}
\end{equation}
with $a_1 = 4.4 \times 10^{-14}\km\s^{-2}$, $a_2= 1.7\times 10^{-14}
\km\s^{-2}$, $z_1=200\p$ and $z_2=1\kpc$.

In this paper we neglect the effects of rotation and velocity shear and excite
an $\alpha^2$-dynamo by imposing an explicit $z$-dependent $\alpha$-effect in
the induction equation,
\begin{equation}
    \label{eq:alpha}
    \alpha(z)=\alpha_0
    \begin{cases}
    \displaystyle
    \sin \left(\pi z/h_\alpha\right)\,, &|z| \leq h_\alpha/2\,,\\
    \displaystyle
    (z/|z|) \exp \left[-\left(2z/h_\alpha-z/|z|\right)^2\right]\,, &|z|>h_\alpha/2\,.
    \end{cases}
\end{equation}

\begin{figure}
    \centering
    \includegraphics[width=0.95\columnwidth]{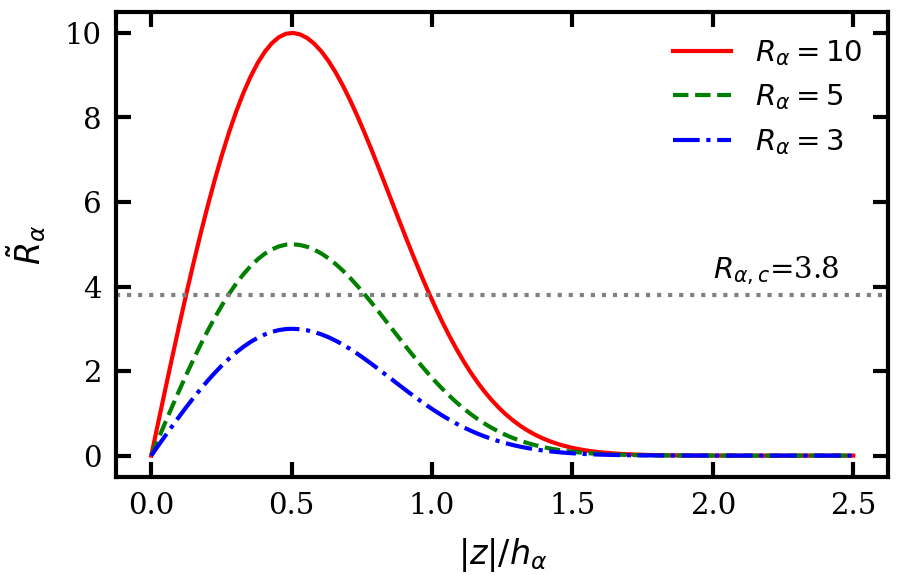}
\caption{
The local strength of the dynamo action,
$\tilde{R}_\alpha=|\alpha(z)|h_\alpha/\eta$ for $\alpha(z)$ of
equation~\eqref{eq:alpha} and various values of $R_\alpha$ of
equation~\eqref{Ral} specified in the legend. The critical value of $R_\alpha$
is shown dotted: the mean-field dynamo is supercritical at those $|z|$ where
$\tilde{R}_\alpha (z)>R_{\alpha,\text{c}}$.}
    \label{alpha_z}
\end{figure}

The vertical extent of the dynamo-active layer is $h_\alpha$ on each side of
the midplane; the smaller is $h_\alpha$, the stronger is the vertical gradient
of the magnetic field and the more it is buoyant. The dynamo intensity (in
particular, the growth rate of the large-scale magnetic field) depends on the
dimensionless number
\begin{equation}\label{Ral}
R_\alpha=\alpha_0 h_\alpha/\eta\,.
\end{equation}
In the Solar neighbourhood of the Milky Way, $\alpha_0\simeq0.5\kms$ and
$h_\alpha\simeq0.5\kpc$ \citep[e.g., p.~317 of][]{SS21}.  We consider a range
of values for these two parameters, as summarised in
Tables~\ref{tab:parameter_values}, \ref{tab:growth_rate_dynamo} and
\ref{tab:growth_rate_dynamo_2}.  The purpose of the form of $\alpha(z)$ adopted
at $|z|>h_\alpha/2$ is to ensure that $|\alpha(z)|$ decreases smoothly from its
maximum $|\alpha|=\alpha_0$ at $|z|=h_\alpha/2$ to negligible values at $|z|\gg
h_\alpha$ since the $\alpha$-effect dynamos are sensitive to discontinuities in
$\nabla\alpha$ which can cause unphysical oscillations of the magnetic field.
Figure~\ref{alpha_z} shows the effective (local) value of $\tilde{R}_\alpha$
defined as in equation~\eqref{Ral} but with $|\alpha(z)|$ of
equation~\eqref{eq:alpha} rather than $\alpha_0$, confirming that the imposed
dynamo action is confined to the layer $|z|\lesssim h_\alpha$ where
$R_\alpha>R_{\alpha,\text{c}}$, and we have determined that
$R_{\alpha,\text{c}}=3.8$ in our case.

The resulting form of the mean-field dynamo is known as the $\alpha^2$-dynamo;
it is weaker than the $\alpha\omega$-dynamo that involves the differential
rotation in galaxies but our goal here is to explore the interaction of the
dynamo and MBI rather than to reproduce in detail the galactic conditions. The
$\alpha^2$-dynamo is a relatively simple and well-understood mean-field dynamo
mechanism; the temporal and spatial scales of the magnetic field produced by
$\alpha^2$-dynamo are known accurately \citep[e.g.,][]{Anvar1983}. This will
help us to distinguish the effects of the dynamo action and the MBI.

The initial conditions represent isothermal hydrostatic equilibrium with the
gas density profile
\begin{equation}
    \label{eq:density_dist}
    \rho(z)=\rho_0\exp\left[\frac{a_1}{{c\sound^2}}\left(z_1-\sqrt{z_1^2+z^2} -\frac{a_2}{2a_1}\frac{z^2}{z_2}\right)\right],
\end{equation}
given a negligible initial magnetic field.  The sound speed $c\sound=15\kms$
corresponds to the temperature $T=2\times10^4\K$ for 50 per cent of hydrogen
ionized, so that the mean molecular weight is 0.74 (assuming 70\% of hydrogen
and 30\% of neutral helium by mass). The resulting initial density scale height
$|\partial\ln\rho/\partial z|^{-1}$ is about $400\p$ at a distance from the
midplane.

The seed magnetic field  applied as an initial condition represents as Gaussian
random noise in the vector potential component $A_z$ with a mean amplitude
proportional to $\rho^{1/2}(z)$ and the maximum strength $10^{-6}\mkG$ at
$z=0$. This field has $B_z=0$.  A random initial magnetic field leads to
shorter transients than in the case of a unidirectional initial field.

\begin{table}
\centering
	\caption{
Simulation runs with the kinematic viscosity $\nu=0.008\kpc\kms$, magnetic
diffusivity $\eta=0.03\kpc\kms$ and various values of $R_\alpha$ of
equation~\eqref{Ral} based on the values of $\alpha_0$ and $h_\alpha$ that
appear in equation~\eqref{eq:alpha}; $\gamma_\text{D}$ is the rate of
exponential growth of the magnetic field strength during the linear phase of
the dynamo, $\gamma_\text{B}$ is the growth rate of the magnetic field obtained
when the magnetic field becomes buoyant and $\gamma_u$ is the growth rate of
the root-mean-square gas speed. A missing entry means that the corresponding
variable does not grow. The last column shows the period $T$ of nonlinear
oscillations of the azimuthal magnetic field (the $y$-component), a missing
entry in this column means that the oscillation period was not measured, in
particular, when it is very long.
}
\label{tab:growth_rate_dynamo}
\adjustbox{width=0.48\textwidth}{
\begin{tabular}{lccccccc}
\hline
Model  &R$_{\alpha}$&$h_{\alpha}$&$\alpha_0$   & $\gamma_\text{D}$ & $\gamma_\text{B}$ & $\gamma_u$ & $T$ \\
       &            & [kpc]      &[km s$^{-1}$]&[Gyr$^{-1}$]&[Gyr$^{-1}$]& [Gyr$^{-1}$] & [Gyr] \\
\hline
\RThA  &          3 & 0.2        & 0.45        & $-$3.8     & --          & --    & --       \\
\RThB  &            & 0.3        & 0.3         & $-$2.5     & --          & --    & --       \\
\RThC  &            & 0.6        & 0.15        &  --        & --          & --    & --       \\[0.2cm]
\RVhA  &          5 & 0.2        & 0.75        & 1.6        & 1.1        &  2.1   &    1.5   \\
\RVhB  &            & 0.3        & 0.5         & 0.4        & --         &  0.6   &    1.9   \\
\RVhC  &            & 0.6        & 0.25        & 0.1        & --         &  --    &    --    \\[0.2cm]
\RHhA  &          7 & 0.2        & 1.05        & 4.4        & 7.9        &  4.21  &    1.4   \\
\RHhB  &            & 0.3        & 0.7         & 1.9        & 2.2        &  2.7   &    1.7   \\
\RHhC  &            & 0.6        & 0.35        & 0.4        & 0.5        &  0.7   &    3.0    \\[0.2cm]
\RXhA  &         10 & 0.2        & 1.5         & 12.4       & 29.1       & 23.8   &    1.5   \\
\RXhB  &            & 0.3        & 1.0         & 5.5        & 11.3       & 7.1    &    1.7   \\
\RXhC  &            & 0.6        & 0.5         & 1.6        & 1.8        & 2.6    &    3.2   \\[0.2cm]
\RXVhA &         15 & 0.2        & 2.25        & 32.2       & 86.9       & 60.3   &    1.6   \\
\RXVhB &            & 0.3        & 1.5         & 14.9       & 36.6       & 27.9   &    1.9   \\
\RXVhC &            & 0.6        & 0.75        &  3.7       & 6.36       & 8.4    &    3.0   \\[0.2cm]
\RXXhA &         20 & 0.2        & 3.0         & 62.7       & 171.4      & 75.9   &    1.6   \\
\RXXhB &            & 0.3        & 2.0         & 28.5       & 74.2       & 49.7   &    1.7   \\
\RXXhC &            & 0.6        & 1.0         &  7.1       & 14.7       & 17.1   &    3.1   \\
\hline
\end{tabular}
}
\end{table}

\section{Mean-field dynamo transformed by magnetic buoyancy}\label{sec:results}

The system that we explore supports both the mean-field dynamo and magnetic
buoyancy instabilities. They can be distinguished clearly only during their
linear stages when the corresponding perturbations grow exponentially. However,
the effects of the dynamo action and magnetic buoyancy are strongly intertwined
at the later stages when the Lorentz force becomes dynamically significant and
the system evolves into a stationary state. It is, correspondingly, harder to
identify the role of each individual instability in the nonlinear behaviour. To
facilitate such an identification, we use a range of models with various
relations between the time scales of the dynamo action and the MBI.

The mean-field dynamo driven by the MBI has been explored earlier (see
references in Section~\ref{sec:intro}) but the effects of the MBI on the dynamo
action driven in spiral galaxies by the turbulence in the gas disc have
received little attention. Our approach is designed to address this aspect of
the evolution of magnetic fields in galaxies and accretion discs. The effects
we find are rather unexpected: the buoyancy of the magnetic fields produced by
the dynamo does not just modify the dynamo but can change entirely the spatial
structure and evolution of the large-scale magnetic field.

\begin{figure}
    \centering
    \includegraphics[width=0.95\columnwidth]{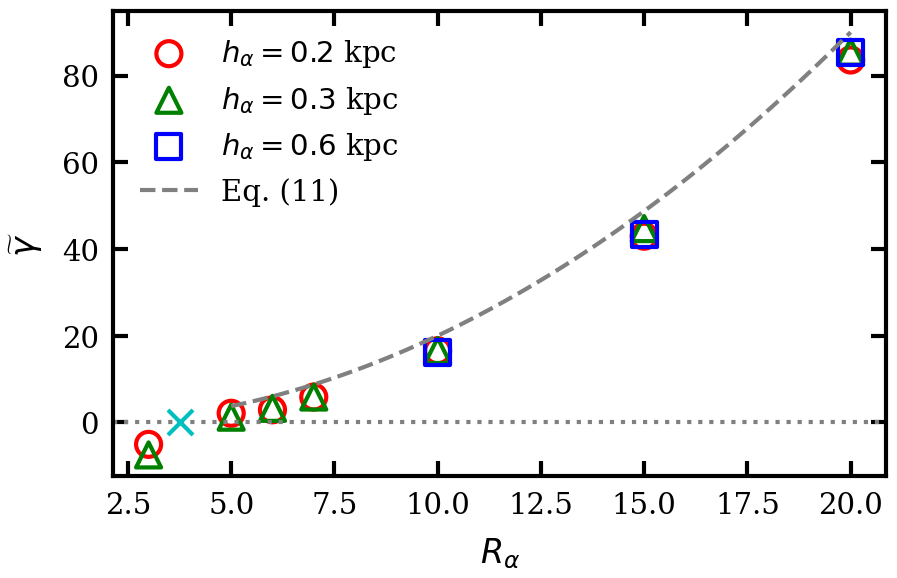}
   \caption{
The dimensionless growth rates $\widetilde{\gamma}=\gamma h_\alpha^2/\eta$ of
the magnetic field at early times of the evolution when the dynamo action is
predominant (symbols, as specified in the legend) and the similarly normalised
dynamo growth rate from equation~\eqref{eq:maximal_growth_rate} for those values of $R_\alpha$ where it applies (solid).
The critical value of $R_\alpha$, corresponding to $\widetilde{\gamma}=0$, is obtained by interpolating the data points as shown with the cross.
  \label{fig:dimensionless_growth_mag}}
\end{figure}

The parameters of the models explored are presented in
Tables~\ref{tab:growth_rate_dynamo} and \ref{tab:growth_rate_dynamo_2}.  Since
the dynamo action is relatively insensitive to the kinematic viscosity $\nu$
while the MBI responds to both viscosity and magnetic diffusivity $\eta$, we
consider systems with $\nu>\eta$ (Table~\ref{tab:growth_rate_dynamo}) and
$\nu<\eta$ (Table~\ref{tab:growth_rate_dynamo_2}).  The results and conclusions
presented below apply in both cases.

The growth rate of the most rapidly growing magnetic field mode under the
$\alpha^2$-dynamo action with $\alpha$ depending on $z$ alone is given by
\citep[Section~4.iii of][]{Anvar1983}
\begin{equation}
    \label{eq:maximal_growth_rate}
\gamma_\text{D} = \frac{\eta}{h_\alpha^2}\left[\tfrac14 R_\alpha^2 -\tfrac12 R_\alpha + O(1)\right]
\end{equation}
for large and moderate values of $R_\alpha$.  As shown in
Fig.~\ref{fig:dimensionless_growth_mag}, the growth rates obtained for the
early stages of the magnetic field growth fit this dependence with high
accuracy. 
The resultant growth rates from equation~\eqref{eq:maximal_growth_rate} are positive when
$R_\alpha>R_{\alpha,\text{c}}=2$. The three-dimensional numerical solutions have a somewhat larger critical value 
$R_{\alpha,\text{c}}\approx3.8$ (obtained by interpolating the data points shown in Fig.~\ref{fig:dimensionless_growth_mag} using the cubic spline for $3\leq R_\alpha\leq7$ ) and smaller $\gamma_\text{D}$ because of the
magnetic diffusion in the $x$- and $y$-directions.  The scale over which the
magnetic field strength in this mode decreases by the factor e is given by
$L_\text{D}\simeq 2h_\alpha R_\alpha^{-1/2}\,,$ and the spatial structure is
oscillatory with the wavelength $\lambda_\text{D}=4\pi h_\alpha/R_\alpha$.  For
$R_\alpha\simeq10$, we have $L_\text{D}\simeq \lambda_\text{D}\simeq h_\alpha$.
The dynamo also supports a wide range of modes at scales larger than about
$L_\text{D}/2$ but they grow at rates lower than that of
equation~\eqref{eq:maximal_growth_rate}. Meanwhile, the scale of the MBI along
the magnetic field is 1--2\,kpc, significantly larger than the values of
$h_\alpha$ that we use. We consider a wide range of $R_\alpha$ to ensure that
the magnetic fields produced by the dynamo and the MBI can have sufficiently
different growth rates and spatial scales to be distinguishable in their linear
stages.

\begin{figure}
    \centering
    \includegraphics[width=0.95\columnwidth]{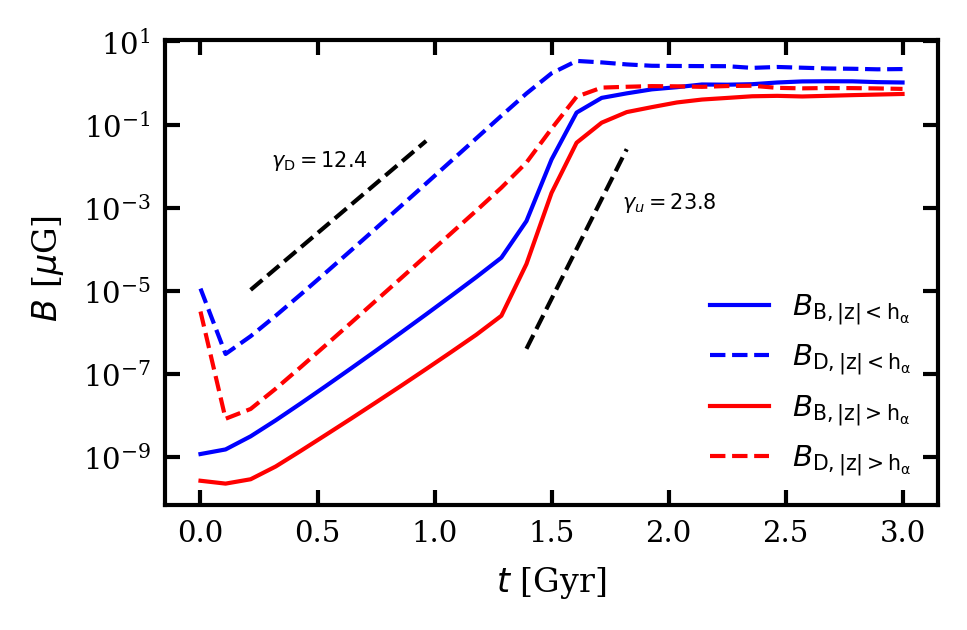}
    \caption{
The evolving strengths of the large- (solid) and small-scale (dashed) magnetic
fields (with the separation scale of $\ell=200\p$) in Model~\RXhA\ averaged
over $|z|<h_\alpha$ (blue) and  $|z|>h_\alpha$ (red).  The dashed lines
represent the exponential growth with the growth rates in Gyr$^{-1}$ from
Table~\ref{tab:growth_rate_dynamo}.
  \label{fig:B_mean_fluc_in_out_midplane_vs_t}
    }
\end{figure}
\begin{figure*}
    \centering
    \includegraphics[width=\textwidth]{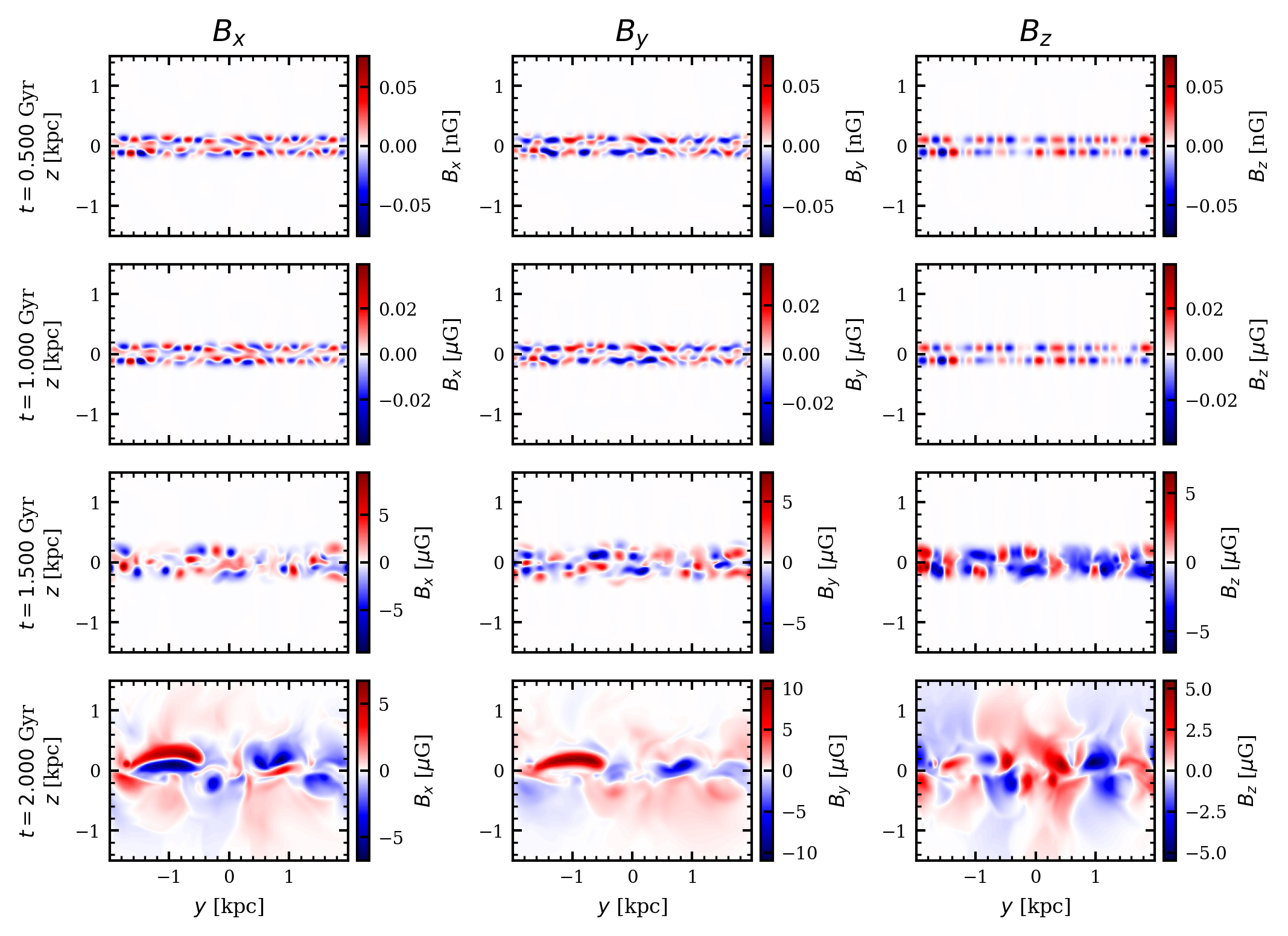}
    \caption{The strength of the horizontally averaged magnetic field components $\meanh{B_x}$, $\meanh{B_y}$ and $\meanh{B_z}$ (columns from left to right) in the $(y,z)$-plane at various evolutionary stages: $t=0.5\Gyr$ 
    and $1\Gyr$ (two upper rows, the linear dynamo phase: the magnetic field strength grows while its spatial structure remains unchanged)
    and $t=1.5$ and $2\Gyr$ (the two bottom rows, an early development of the MBI and the nonlinear stage) in Model~\RXhA.
    }
    \label{fig:components_of_B_in_time}
\end{figure*}

Our simulations start with a weak random magnetic field that launches the
dynamo action. The growth rate of the MBI is of the order of
$V_\text{A}/h_\alpha$ (with  $V_\text{A}$ the Alfv\'en speed), so there is a
relatively long period when the magnetic field is too weak to be
unstable because of its buoyancy. The growth rate of the MBI exceeds that of
the dynamo when $V_\text{A}\gtrsim\alpha_0/4$; this happens when the magnetic
field is relatively strong. Depending on the system parameters, the second
phase of the system evolution when the magnetic field generated by the dynamo
becomes unstable with respect to the MBI may or may not be clearly discernible.

Figure~\ref{fig:B_mean_fluc_in_out_midplane_vs_t} presents Model~\RXhA, in which
the growth rates and characteristic scales of the mean-field dynamo and MBI
differ noticeably. $L_\text{D}\simeq0.13\kpc$ for the scale at which magnetic
field strength decreases and $\lambda_\text{D}\simeq0.25\kpc$ for the
wavelength at which the field direction changes.  The evolving spatial
structure of the magnetic field is illustrated in
Fig.~\ref{fig:components_of_B_in_time}. At early times after the short-term
transients from the initial condition have dissipated, the magnetic field is
weak, has a relatively small scale imposed by the dynamo and is confined to a
thin disc $|z|\lesssim h_\alpha$ where the $\alpha$-effect is imposed.  At a
later time, $t=1.4\Gyr$, the magnetic field spreads out of the disc because of
the magnetic diffusion and the first signs of the magnetic buoyancy emerge,
particularly manifested in the enhancement of $B_z$ (the right-hand column of
Fig.~\ref{fig:components_of_B_in_time}). At that time, the growth rate of the
magnetic field increases (Fig.~\ref{fig:B_mean_fluc_in_out_midplane_vs_t}). The
effect of the magnetic buoyancy becomes even more significant at $t=1.5\Gyr$ --
at this stage, the exponential field amplification typical of a linear
instability continues at the enhanced rate. In the strongly nonlinear stages at
$t>1.5\Gyr$, the magnetic field structure characteristic of the mean-field
dynamo is completely wiped away despite the continued action of the
$\alpha$-effect.

\begin{figure*}
    \centering
    \includegraphics[width=\textwidth]{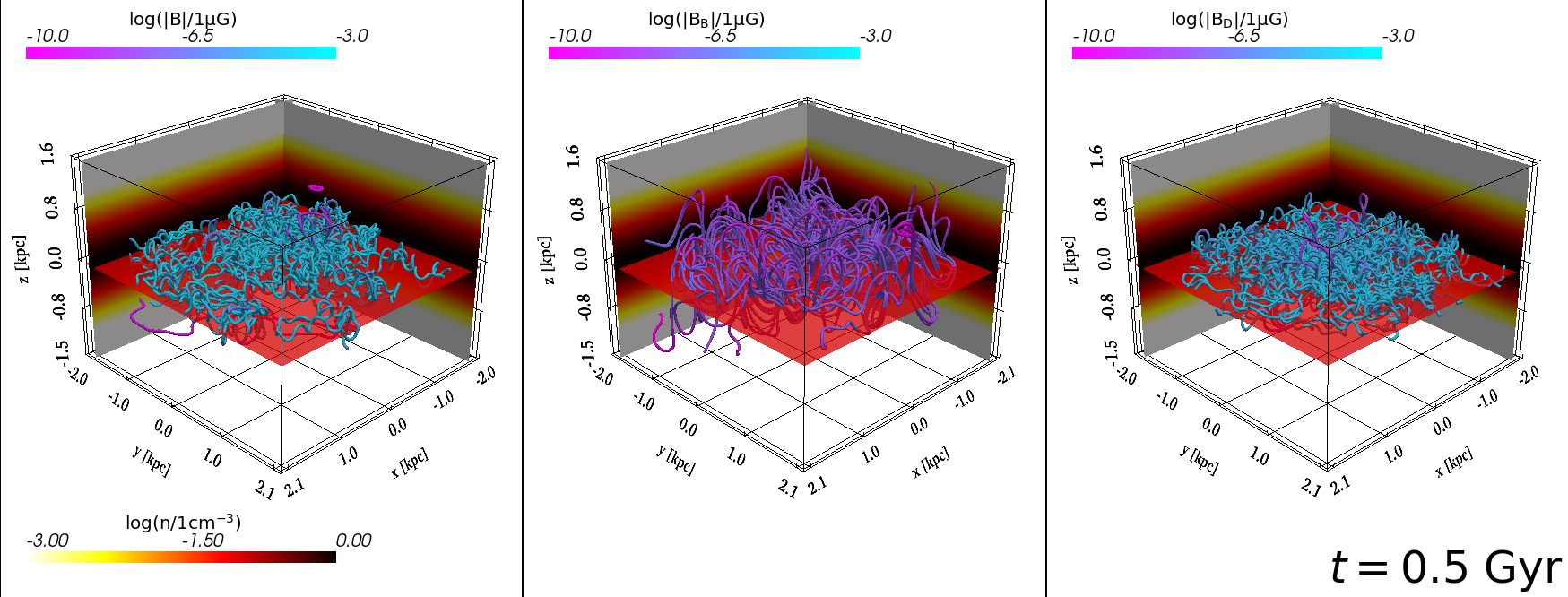}
    \includegraphics[width=\textwidth]{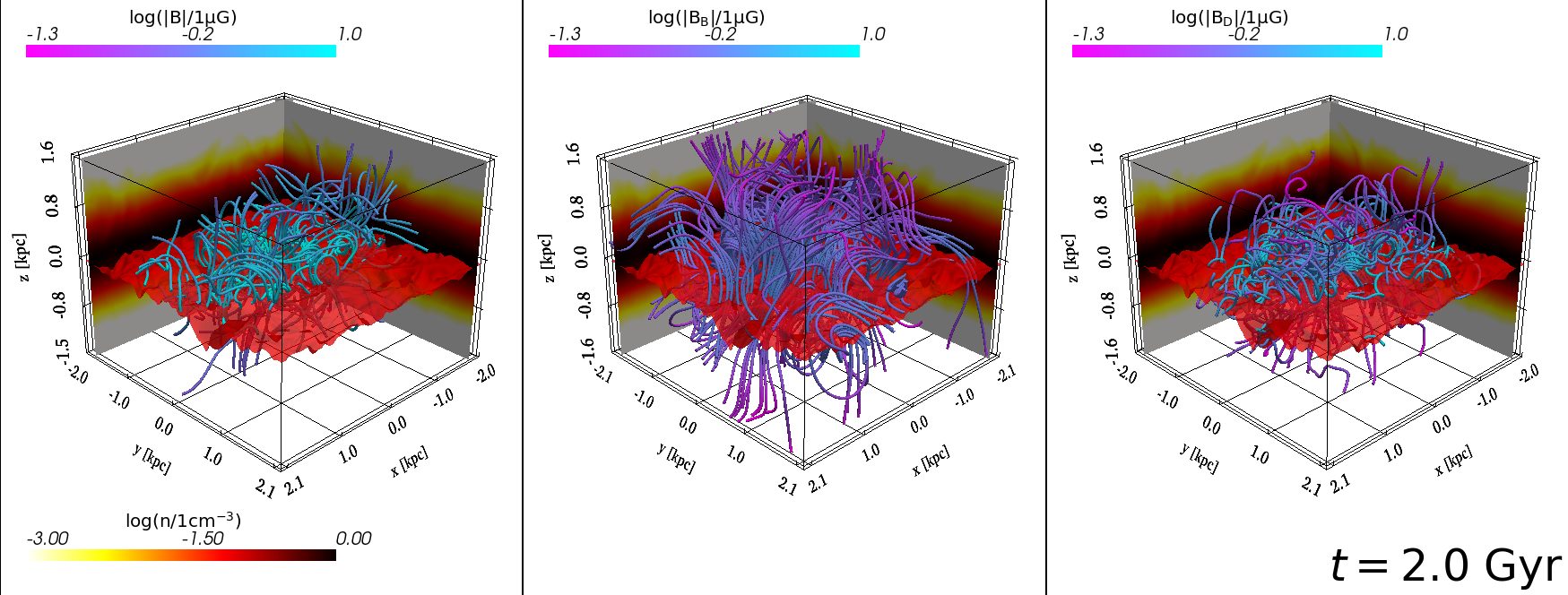}
    \caption{
The magnetic lines in Model~\RXhA\ of the total field $\vec{B}$ (left-hand
column), separated using equation~\eqref{GS} with $\ell=200\p$ into
contributions at the larger scales characteristic of the magnetic buoyancy
$\vec{B}_\text{B}$ (middle) and the smaller scales of the imposed dynamo
$\vec{B}_\text{D}$ (right-hand column). The red isosurface corresponds to
a gas number density of $0.7{\cm}^{-3}$.
    \label{fig:streamlines}
}
\end{figure*}

\begin{figure*}
    \centering
master\includegraphics[width=\textwidth]{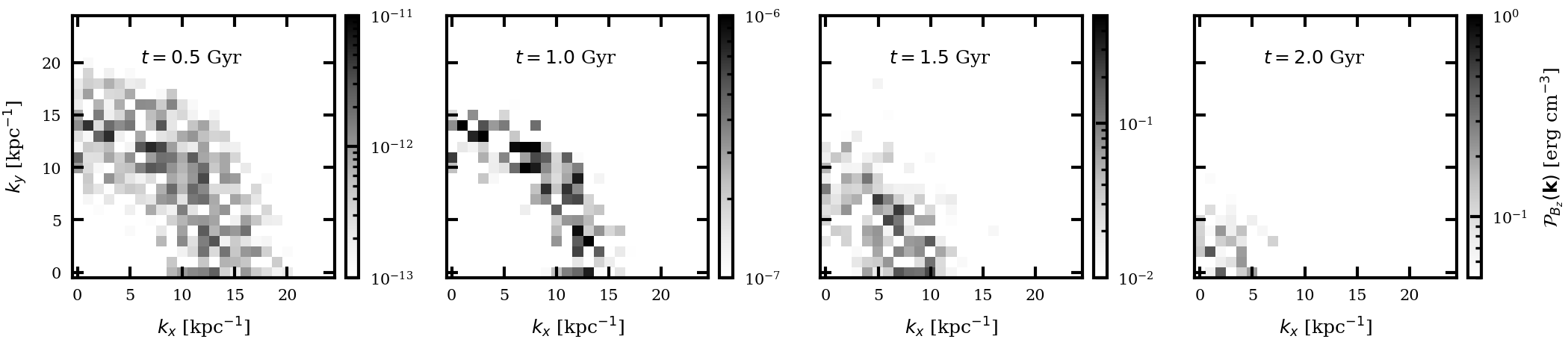} 
    \caption{
Two-dimensional power spectra in the $(k_x,k_y)$-plane of $B_z$ in Model~\RXhA\
(the same as in Fig.~\ref{fig:B_mean_fluc_in_out_midplane_vs_t}) at $z=385\p$
during the dominance of the kinematic dynamo action (left), two transitional
phases (middle) and a stationary state (right).
    \label{fig:power_spectrum_dynamo}
}

\end{figure*}

\label{sec:dynamo-spectra}

To illustrate the three-dimensional magnetic field structure, including the
magnetic loops produced by the MBI and the magnetic field generated by the
dynamo transformed by the MBI, Fig.~\ref{fig:streamlines} shows the evolving
three-dimensional structure of magnetic lines at various scales. The total
magnetic field $\vec{B}$ in this figure and in
Fig.~\ref{fig:B_mean_fluc_in_out_midplane_vs_t} is separated into the
contributions $\vec{B}_\text{B}$ of the larger scales (characteristic of the
MBI) and of the smaller ones $\vec{B}_\text{D}$ (driven by the imposed dynamo
action) using the Gaussian smoothing,
\begin{equation}\label{GS}
\vec{B}_\text{B}(\vec{x},t)= \int_V \vec{B}(\vec{x}',t)\, G_\ell(\vec{x}-\vec{x}')\,\dd^3\vec{x}'\,,
\quad
\vec{B}_\text{D}=\vec{B}-\vec{B}_\text{B}\,,
\end{equation}
where the integration extends over the whole domain volume with the smoothing
kernel  $G_\ell(\vec{\xi})=(2\pi\ell^2)^{-3/2}\exp[-|\vec{\xi}|^2/(2\ell^2)]$
with $\ell=200\p$.
The smoothing scale is chosen to be equal to $L_D\simeq\lambda_D\simeq h_\alpha$, the scale of the spatial variations of the magnetic field generated by the dynamo as estimated above, and $h_\alpha=200\p$ in Fig.~\ref{fig:streamlines}.
We note that the small-scale part $\vec{B}_\text{D}$ also
contains random magnetic fields produced by nonlinear effects at the later
stages of the system's evolution.

\begin{figure}
    \centering
\includegraphics[width=0.95\columnwidth]{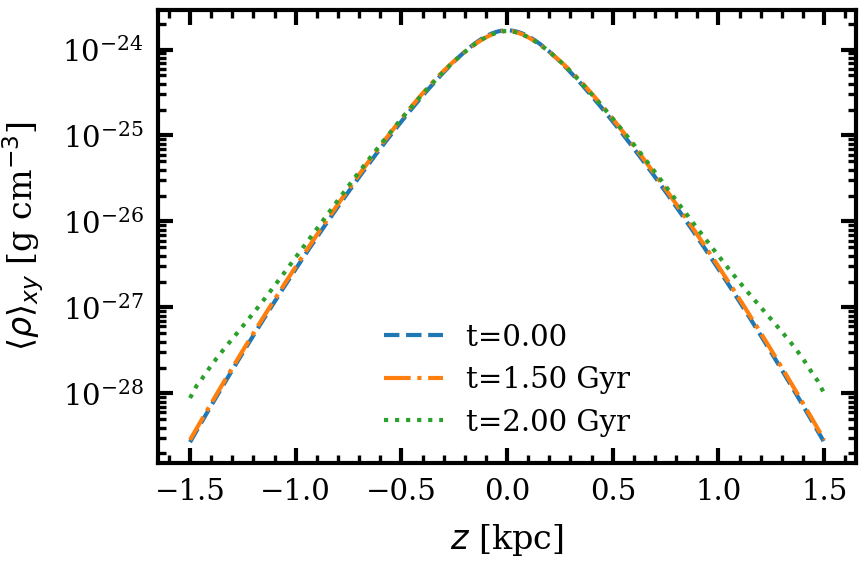}
    \caption{The horizontal average of the gas density in the Model~\RXhA, in
the initial condition ($t=0$), in the early MBI phase ($t=1.5\Gyr$) and in the
nonlinear state ($t=2\Gyr$).
    \label{fig:rhomz_R10h2}
}
\end{figure}

The restructuring of the magnetic field by the MBI is quantified in
Fig.~\ref{fig:power_spectrum_dynamo}, which shows the two-dimensional power
spectra of the $z$-component of the magnetic field at the times indicated above
each frame.  These confirm the evolution pattern visible in
Fig.~\ref{fig:components_of_B_in_time}.  Over time the dominant horizontal
scales $2\pi k_x^{-1}$ and $2\pi k_y^{-1}$ of the magnetic field grow larger:
at $t\lesssim0.5\Gyr$, the spectrum centres around $k_y \approx k_x \approx
2\pi/\lambda_\text{D}\approx 15\kpc^{-1}$, which reduces by $t=2\Gyr$ to $k_x
\simeq k_y\lesssim 5 \kpc^{-1}$.  The latter wavenumbers correspond to scales
in excess of 1--$2\kpc$ characteristic of the MBI. As the peak wavenumbers
decrease their spread is broader as the MBI excites a wider range of unstable
modes. Typical of the $\alpha^2$-dynamo, the magnetic field components $B_x$
and $B_y$ are of about equal strength and scale (velocity shear due to
differential rotation would enhance the azimuthal component $B_y$ in comparison
with $B_x$ and modify the field scales to $k_y<k_x$).

Figure~\ref{fig:rhomz_R10h2} shows that the  distribution in $z$ of the
horizontally averaged gas density varies very little with time, only becoming
slightly wider in the non-linear phase as the magnetic buoyancy carries matter
away from the midplane. This behaviour is quite different than in the
simulations of the MBI and Parker instability of an \textit{imposed} magnetic
field. There the scale height of the magnetic field (and cosmic rays) increases
in the nonlinear state so strongly that the gas layer loses the magnetic and
cosmic-ray pressure support as the system evolves and becomes much thinner, as
shown in Fig.~11d of \citet{DT2022a} and Fig.~4 of \citet{DT2022b}.  
Remarkably and unexpectedly, the nonlinear states of the MBI (and the Parker instability) are sensitive to the mechanism by which the magnetic field is maintained (imposed versus mean-field dynamo in our case).

\section{Nonlinear oscillations}\label{sec:nonlinear-oscilations}

\begin{figure}
    \centering
\includegraphics[width=\columnwidth]{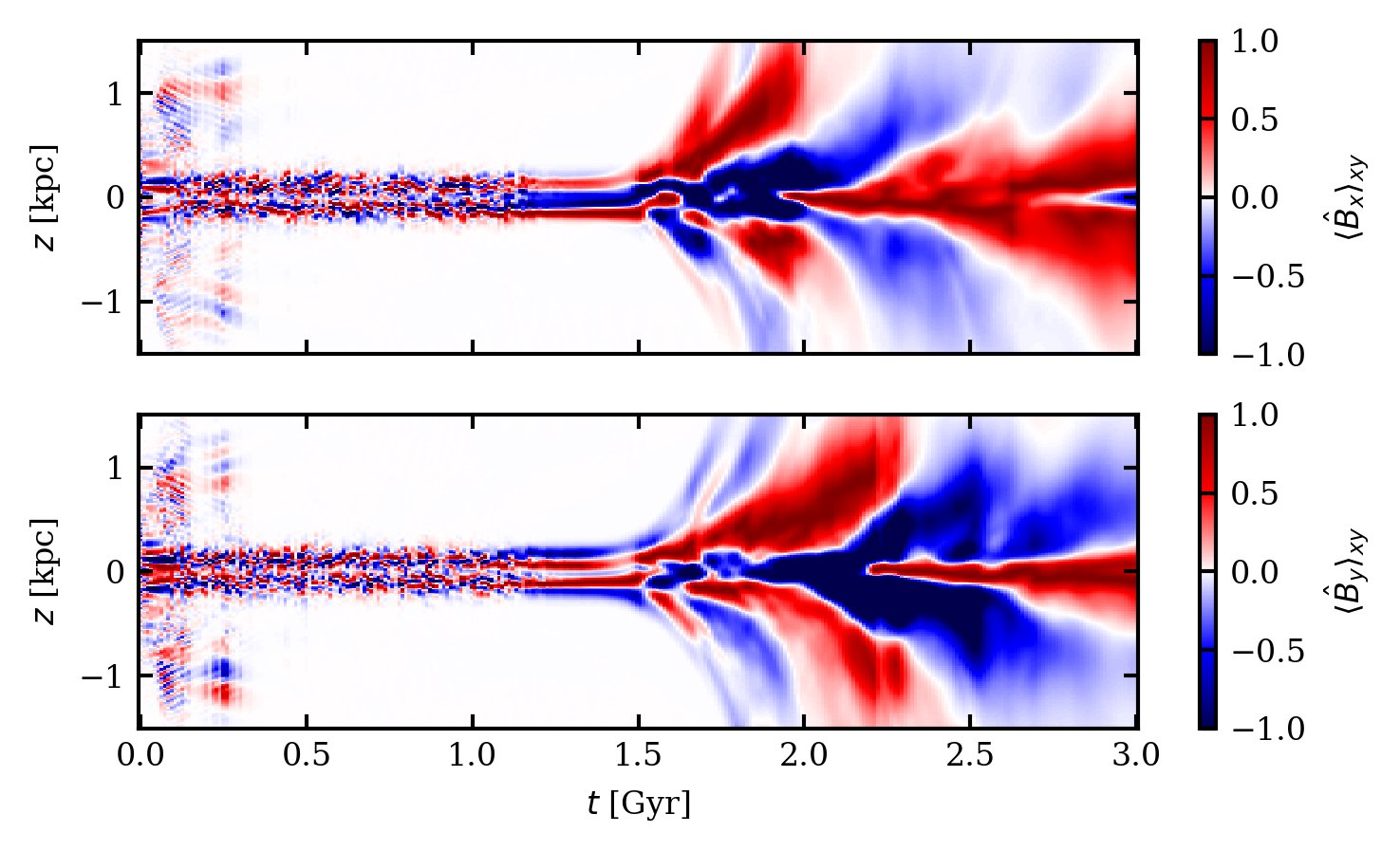}
    \caption{The evolution of the horizontally averaged magnetic field components $\meanh{\widehat{B}_x}$ (upper panel) and $\meanh{\widehat{B}_y}$ (lower panel) in Model~\RXhA\ normalised to their maximum values at each time.
    \label{fig:xy_Bx_By_ha_02_ra_10}
}
\end{figure}

\begin{figure}
    \centering
\includegraphics[width=0.95\columnwidth]{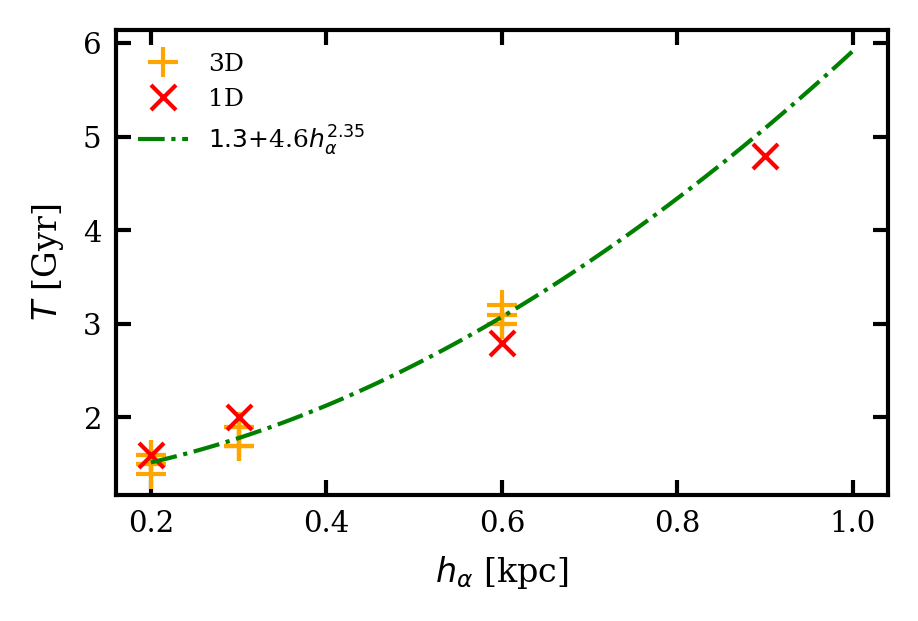}
    \caption{Variation of the period of the nonlinear oscillations of the
$y$-component of the magnetic field  $h_{\alpha}$: from three-dimensional
simulations (pluses, also presented in  Table~\ref{tab:growth_rate_dynamo}) and
the one-dimensional model of Section~\ref{sec:1D} with $\alpha_0 = 1.5\kms$
(crosses). The fit to the three-dimensional simulation results is shown
dash-dotted its form is specified in the legend with $h_\alpha$ in kiloparsecs.
    \label{fig:T_vs_ha_1D}
}
\end{figure}

Figure~\ref{fig:xy_Bx_By_ha_02_ra_10} illustrates an even more fundamental
consequence of the MBI: the magnetic field, which grows monotonically at early
times, develops oscillations at $t\gtrsim1.5\Gyr$ when it becomes strong enough
to make the system essentially nonlinear. The figure shows the evolution of the
horizontally averaged magnetic field components $\meanh{B_x}$ and $\meanh{B_y}$
from Model~\RXhA, normalised to their maximum values at each time to better
expose the field structure at early times when it is still weak.

The magnetic field generated by the linear dynamo is confined to a relatively
thin layer $|z|\lesssim h_\alpha$ and grows monotonically. However, it spreads
to larger altitudes because of the buoyancy (to achieve the scale height of
order $1\kpc$). When fully nonlinear the magnetic field becomes oscillatory,
reversing its direction at intervals of order $0.5\Gyr$.  There is a phase
shift in Model~\RXhA of around a quarter of a period between the $x$- and
$y$-components of the magnetic field.  The period of the oscillations is
presented in the last column of Table~\ref{tab:growth_rate_dynamo}.

The field scale height, the timing of the onset of the oscillations and their
period vary with $h_\alpha$, but the oscillations are a general property of the
models with sufficiently strong magnetic fields.

Figure~\ref{fig:T_vs_ha_1D} shows the variation of the oscillation period with
$h_{\alpha}$ in both the three-dimensional simulations presented in
Table~\ref{tab:growth_rate_dynamo} and the 1D model of Section~\ref{sec:1D} for
$\alpha_{0}=1.5 \kms$.  Table~\ref{tab:growth_rate_dynamo} shows that the
period $T$ does not exhibit any significant variation with $\alpha_0$ for a
fixed $h_\alpha$, but does change systematically with $h_\alpha$ for a fixed
$R_\alpha$, the dimensionless measure of the kinematic dynamo intensity.  This
is understandable since the oscillations occur in the nonlinear stage which is
not very sensitive to the strength of the \textit{kinematic} dynamo.  The
increase of the oscillation period with $h_\alpha$ can be attributed to the
change in the characteristic time of the system, the sound (or Alfv\'en)
crossing time $h_\alpha/c\sound$ if the oscillations are driven by the magnetic
buoyancy or the magnetic diffusion time $h_\alpha^2/\eta$ if they are a
nonlinear dynamo phenomenon.  When normalised to the magnetic diffusion time
$h_\alpha^2/\eta$, the oscillation period can be approximated for all the
models explored (either three- or one-dimensional) as (shown dash-dotted in
Fig.~\ref{fig:T_vs_ha_1D})
\begin{equation}\label{T}
\left(\frac{T}{1\Gyr}-1.3 \right)\frac{\eta}{h_\alpha^2} \approx
0.14\left(\frac{h_\alpha}{1\kpc} \right)^{0.35}\,.
\end{equation}
The residual variation, $h_\alpha^{0.35}$, is weak enough to suggest that the
oscillations are mainly driven by the dynamo action triggered by the magnetic
buoyancy as discussed in Section~\ref{HFDBMF}.

Magnetic field reversals, which are likely to be a manifestation of similar
nonlinear oscillations, also occur in the simulations of the MBI of an imposed
magnetic field \citep{DT2022b}.  \citet{JoLe08} and \citet{GJL12} also observe
a reversal of the large-scale magnetic field in their simulations of the
magnetic buoyancy in accretion discs. As we argue in Section~\ref{HFDBMF}, they
are produced by the mean-field dynamo action at $|z|\gtrsim h_\alpha$
associated with the helical gas flows driven by the Lorentz force.

\subsection{One-dimensional model of nonlinear oscillations}\label{sec:1D}\label{sec:mf}
The kinematic stage of the $\alpha^2$-dynamo at $t\lesssim1.3\Gyr$ in
Model~\RXhA\ is non-oscillatory. Indeed, the kinematic $\alpha^2$-dynamo is
known to develop an oscillatory solution only of a long period
\citep{Anvar1983,ShSoRu85,RaBr87} or because of the boundary effects
\citep{BaSh87}. In this section, we propose a nonlinear one-dimensional model
of the $\alpha^2$-dynamo with advection due to the magnetic buoyancy and
demonstrate that it not only admits oscillatory solutions but reproduces
quantitatively and in fine detail the oscillatory behavior of the magnetic
field discussed above.

The $x$- and $y$-components of the mean-field dynamo equation
\begin{equation}\label{na2}
\deriv{\mean{\vec{B}}}{t} = \nabla\times(\vec{U}\times\mean{\vec{B}})+
\nabla\times(\alpha\mean{\vec{B}})+\beta\nabla^2\mean{\vec{B}}
\end{equation}
for the mean magnetic field $\mean{\vec{B}}$  and the mean velocity
$\mean{\vec{u}}\equiv \vec{U}=(0,0,U_z)$ can be written as
\begin{align}\label{MFDx}
\deriv{B_x}{t}  + U_z\deriv{B_x}{z} &= -\deriv{}{z}(\alpha B_y) -B_x\deriv{U_z}{z} +\beta\deriv{^2 B_x}{z^2}\,, \\
\label{MFDy}
\deriv{B_y}{t} + U_z\deriv{B_y}{z} &= \deriv{}{z}(\alpha B_x) -B_y\deriv{U_z}{z}  +
            \beta\deriv{^2 B_y}{z^2}\,,
\end{align}
where we assume that all variables only depend on $t$ and $z$ (the infinite
slab approximation) and we have omitted brackets denoting averaging to simplify
the notation in this section. As follows from $\nabla\cdot\vec{B}=0$,
$B_z=\text{const}$ in this approximation. The advection velocity $U_z$
satisfies the Navier--Stokes equation
\begin{equation}\label{eq:mmom}
\deriv{U_z}{t} +U_z\deriv{U_z}{z}=
- \frac{1}{\rho_0}\nabla\left(\frac{|\vec{B}|^2}{8\pi} +c\sound^2\nabla\rho_0 \right) + \frac{\rho'}{\rho_0}g + \nu\deriv{^2 U_z}{z^2}\,,
\end{equation}
where $g$ is the vertical acceleration due to gravity and the second term on
the right-hand side is the Archimedes force resulting from magnetic buoyancy.
We neglect the time variation of the gas density, adopting $\rho=\rho_0$ at all
times but, in the spirit of the Boussinesq approximation, include density
variation $\rho'$ in the Archimedes force. Consider a region of density
$\rho=\rho_0+\rho'$ containing a magnetic field of a strength $B+b$ surrounded
by the gas of the density $\rho_0$ with magnetic field $B$ (here $B$ is the
mean field strength and $b$ is its local perturbation). The pressure balance in
an isothermal gas then leads to
\begin{equation}\label{drho}
\rho'=-\frac{2Bb+b^2}{8\pi c\sound^2}\,.
\end{equation}

Equations~\eqref{MFDx}--\eqref{eq:mmom} are solved numerically in $0< z < z_0$
with $z_0=1.5\kpc$.  $U_z$, $B_x$ and $B_y$ are symmetric about $z=0$ (a
quadrupolar magnetic structure, evident in Fig.~\ref{fig:xy_Bx_By_ha_02_ra_10}
and known to dominate in a thin layer -- e.g., Section~11.3.1 of
\citealt{SS21}) with
\begin{equation}\label{bc1d1}
\deriv{B_x}{z} = \deriv{B_y}{z} = U_z = 0.
\end{equation}
At $z=z_0$ we apply an impenetrable boundary condition for $U_z$, and  vacuum
boundary conditions for the magnetic field with
\begin{equation}\label{bc1d2}
U_z=B_x = B_y = 0,
\end{equation}
justified by the fact that the turbulent magnetic diffusivity increases with
$|z|$ \citep[see Section 11.3 of][for details]{SS21}. 
Larger vertical sizes were used to confirm the domain was large enough to prevent any spurious
boundary effects over the simulation period.  We use the form of $\alpha$ given
in equation~\eqref{eq:alpha} and an initial magnetic field of about
$10^{-6}\mkG$ in strength.

\begin{figure}
    \centering
    \includegraphics[width=0.95\columnwidth]{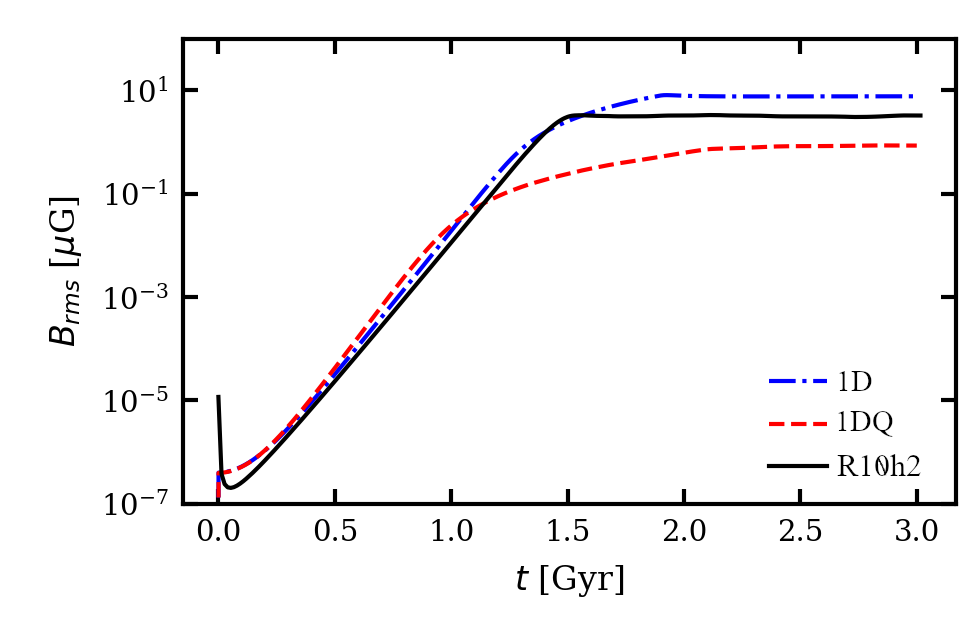}
    \caption{
The evolution of the magnetic field strength in the Model~\RXhA\ (solid/black)
and the one-dimensional mean-field model of Section~\ref{sec:1D} without
(dash-dotted/blue) and with (dashed/red) $\alpha$-quenching of
equation~\eqref{alpq} with $B_0=3\mkG$ obtained using the same values of all
relevant parameters.
\label{fig:quenching_stat_level}
}
\end{figure}

Equations~\eqref{MFDx}--\eqref{eq:mmom} are nonlinear and the dynamo action can
saturate due to the magnetic buoyancy alone.
Figure~\ref{fig:quenching_stat_level} compares the root-mean-square strengths
of the magnetic field in the Model~\RXhA\ (solid, the reference model
discussed above), and in the one-dimensional model with the same values of
$R_\alpha$ and $h_\alpha$ (dash-dotted): the agreement in the growth rate of
the magnetic field and the development of its steady state is quite
satisfactory.

However, magnetic buoyancy is not the only nonlinearity and, as discussed in
Section~\ref{HFDBMF}, the growing magnetic field also suppresses the
$\alpha$-effect via the magnetic current helicity $\alpha\magn$. Therefore, we
also considered a modification of the one-dimensional model that includes the
quenching of the $\alpha$-effect by the magnetic field, a widely used heuristic
form of the dynamo nonlinearity where $\alpha$ in
equations~\eqref{MFDx}--\eqref{MFDy} is replaced with
\begin{equation}\label{alpq}
\frac{\alpha}{1+(B/B_0)^2}\,,
\end{equation}
where $B=(B_x^2+B_y^2)^{1/2}$ and $B_0=3\mkG$.  Apart from modifying the
transition to the steady state and the magnetic field strength at which the
dynamo action saturates, both evident in Fig.~\ref{fig:quenching_stat_level},
the additional nonlinearity affects the period of the magnetic field
oscillations, confirming once again the nonlinear nature of the oscillatory
behaviour.

\begin{figure}
\centering
\includegraphics[trim=0.2cm 0.2cm 0.2cm 0.1cm,clip=true,width=\columnwidth]{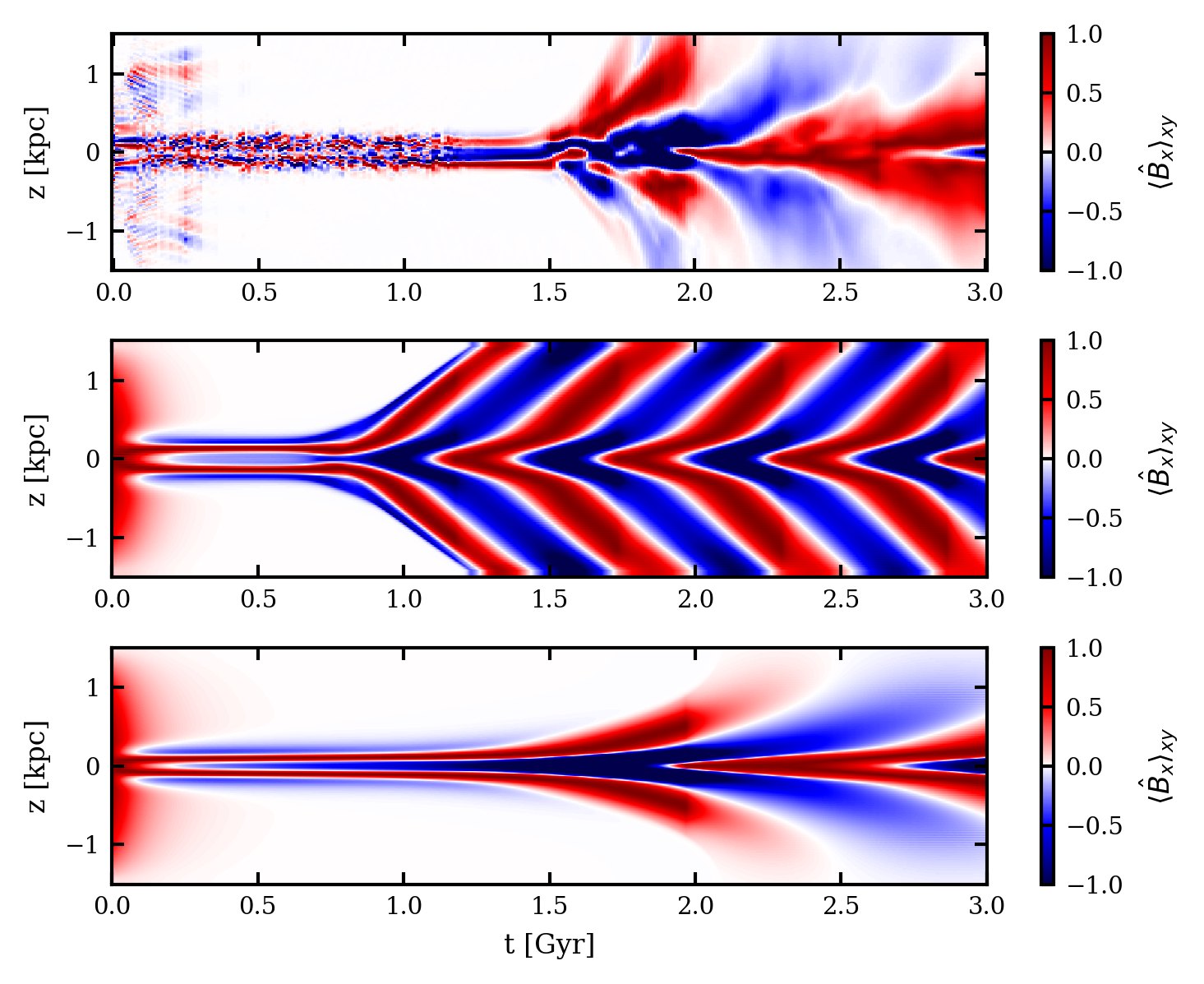}
\caption{The evolution of the horizontally averaged $x$-component of the
magnetic field in Model~\RXhA\ (top panel, identical to the upper panel of
Fig.~\ref{fig:xy_Bx_By_ha_02_ra_10}) and in the one-dimensional model of
Section~\ref{sec:1D} without (middle) and with (bottom) $\alpha$-quenching of
equation~\eqref{alpq} obtained using the common parameters $R_\alpha=10$ and
$h_\alpha=0.2\kpc$. Magnetic field strengths are normalised to unity at each
time.
}
\label{fig:DNS_vs_RK_vs_RKQ_Bx}
\end{figure}

\begin{figure}
\centering
\includegraphics[trim=0.2cm 0.2cm 0.2cm 0.1cm,clip=true,width=\columnwidth]{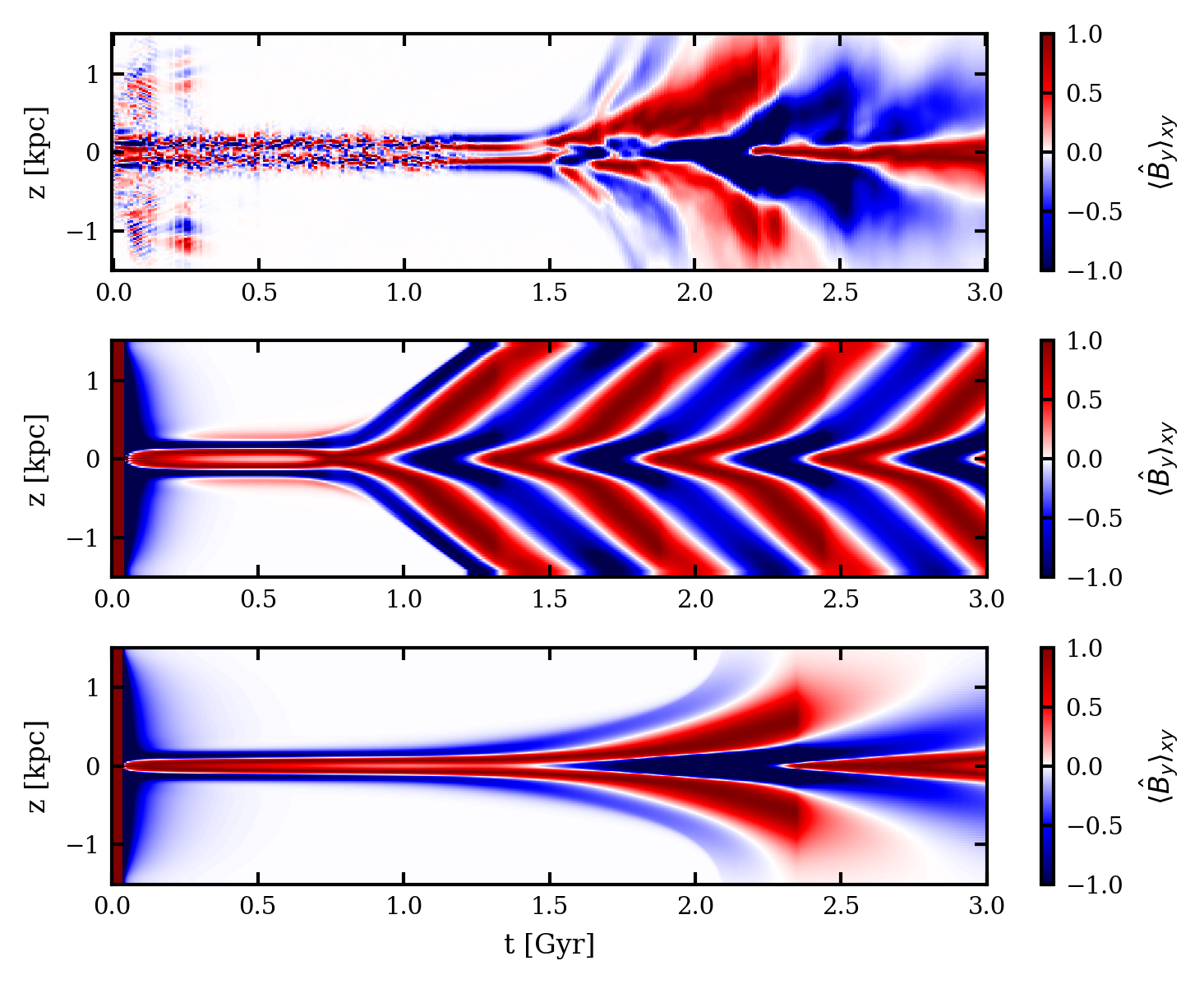}
\caption{As in Fig.~\ref{fig:DNS_vs_RK_vs_RKQ_Bx} but for the $y$-component of
the magnetic field.}
\label{fig:DNS_vs_RK_vs_RKQ_By}
\end{figure}

Figures~\ref{fig:DNS_vs_RK_vs_RKQ_Bx} and \ref{fig:DNS_vs_RK_vs_RKQ_By} compare
the magnetic field evolution, including its oscillations obtained in
Model~\RXhA\ with those from the one-dimensional model. Solution of
equations~\eqref{MFDx}--\eqref{eq:mmom} with $\alpha$ a function of $z$ alone
and independent of $B$ does oscillate but the period of the oscillations is
shorter than in the three-dimensional simulations. The additional
$\alpha$-quenching of equation~\eqref{alpq} makes the oscillation period longer
and dependent on $B_0$. In the lower panels, we have adjusted $B_0$ to obtain
the oscillations at about the same period as in the three-dimensional
simulation although this affects the saturation level of the magnetic field.

We do not attempt to achieve a precise match between the three-dimensional and
one-dimensional results being content with the fact that the one-dimensional
model justifies further our conclusion that the magnetic oscillations are an
essentially nonlinear phenomenon that relies on an interaction of the
mean-field dynamo and magnetic buoyancy not envisaged earlier.

\begin{figure}
    \centering
    \includegraphics[trim=0.2cm 0.2cm 0.2cm 0.1cm,clip=true,width=\columnwidth]{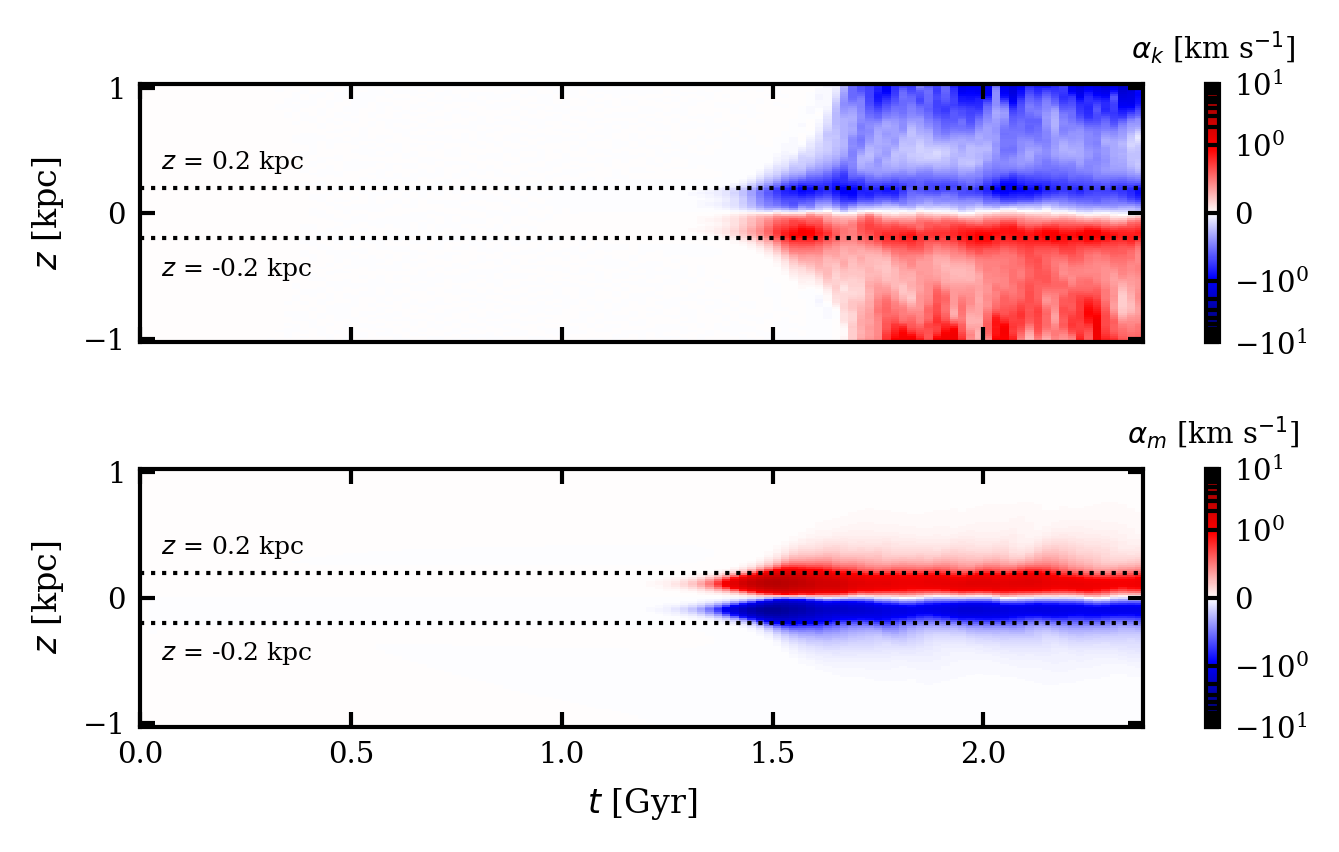}\\
    \caption{
The evolution of the horizontally averaged mean kinetic helicity (upper panel)
and mean magnetic helicity (lower panel) coefficients of the random velocity
and magnetic fields, equations~\eqref{alphak} and \eqref{alpham} respectively,
in Model~\RXhA. The horizontal dotted lines are shown at $|z|=h_\alpha$.
}
\label{fig:alpha-effects}
\end{figure}

\section{Helical flows and dynamo driven by a buoyant magnetic field}\label{HFDBMF}
Figure~\ref{fig:alpha-effects} shows the mean kinetic helicity coefficient of
the gas flow (this does not include the imposed $\alpha$-effect) defined as
\begin{equation}\label{alphak}
\alpha\kin=-\tfrac13 \tau\mean{\vec{u}\cdot(\nabla\times\vec{u})}\,,
\end{equation}
computed using the gas velocity $\vec{u}$ in the simulations and the flow
correlation time $\tau$ obtained using the  autocorrelation function of
$|\vec{u}|$ as explained in Appendix~\ref{appendix:cor_time_len}.
Figure~\ref{fig:alpha-effects} shows $\alpha\kin$ additionally averaged over
$x$ and $y$ (the horizontal average), $\meanh{\alpha\kin}$. Apart from the
$\alpha$-effect, large-scale magnetic fields are subject to the turbulent
magnetic diffusivity with the coefficient
\begin{equation}\label{tdiff}
\beta=\tfrac13\tau\mean{u^2}\,.
\end{equation}
Since Model~\RXhA\ does not include rotation, the mean helicity of the flow
$\mean{\vec{u}\cdot(\nabla\times\vec{u})}$ can only be driven by the Lorentz
force. The Coriolis force in a stratified, rotating system is the cause of the
conventional $\alpha$-effect with $\alpha\kin>0$ for $z>0$ and
$\alpha\kin(-z)=-\alpha\kin(z)$ \citep[e.g., Section 7.1 of][]{SS21}. While the
antisymmetry of $\alpha\kin$ in $z$ is evident in Fig.~\ref{fig:alpha-effects},
the sign of $\alpha\kin$ is opposite to that of the $\alpha$-effect produced by
the Coriolis force. The mean-field dynamo action imposed near the midplane,
with $\alpha$ as given in equation~\eqref{eq:alpha}, has the conventional sign,
$\alpha>0$ at $z>0$. However, the magnetic field generated by the dynamo drives
helical motions that have the opposite sign of the mean helicity, which leads
to the saturation of the dynamo action in the layer near the midplane where the
$\alpha$-effect is imposed \citep[e.g., Section~7.11 of][]{SS21}. Because of
the magnetic buoyancy, the magnetic field spreads out from the layer $|z|\leq
h_\alpha$ and drives motions with the `anomalous' mean helicity, $\alpha\kin<0$
at $z>0$ and $\alpha\kin>0$ at $z<0$ producing the picture shown in
Fig.~\ref{fig:alpha-effects}. Indeed, the gas flow becomes helical only at
later times $t\ga 1.5\Gyr$ when the magnetic field has become strong enough and
has spread to large $|z|$. Since the system is highly nonlinear, significant
parts of the velocity and magnetic fields are randomised as shown in
Figs.~\ref{fig:components_of_B_in_time} and \ref{fig:streamlines}.

\citet[][their Fig.~13]{DT2022b} similarly find that the sign of the mean
kinetic helicity is anomalous, with $\alpha\kin<0$ at $z>0$, in their
simulations of the magnetic buoyancy of an imposed magnetic field in a rotating
system.

Thus, the mean magnetic field generated near the midplane spreads to larger
altitudes where it drives helical flows and an ensuing mean-field dynamo, but
the sign of the $\alpha$-effect is opposite to the sign of the imposed
$\alpha$-effect near the midplane. As we show in Section~\ref{sec:1D}, the
resulting complex dynamo system generates oscillatory magnetic fields.

As with any other dynamo action, the dynamo driven by the Lorentz force away
from the midplane saturates by reducing the magnitude of the $\alpha$-effect,
so that the total $\alpha$-coefficient is modified as
$\alpha=\alpha\kin+\alpha\magn$, where \citep[Section 7.11 of][]{SS21}
\begin{equation}\label{alpham}
\alpha\magn=\frac{\tau}{3}\,\frac{\mean{\vec{b}\cdot(\nabla\times\vec{b})}}{4\pi\rho}\,,
\end{equation}
where $\vec{b}=\vec{B}-\mean{\vec{B}}$ is the deviation of the total magnetic
field $\vec{B}$ from its mean $\mean{\vec{B}}$. To compute $\vec{b}$ and
then $\alpha\magn$, we derived $\mean{\vec{B}}$ by smoothing the total field
with a Gaussian kernel as in equation~\eqref{GS} with a smoothing length of
$\ell=200\p$.  As expected, $\alpha\magn$, shown in the lower panel of
Fig.~\ref{fig:alpha-effects}, has the sign opposite to that of $\alpha\kin$ and
a comparable magnitude.

\begin{figure*}
    \centering
\includegraphics[trim=0.2cm 0.2cm 0.2cm 0.1cm,clip=true,width=\textwidth]{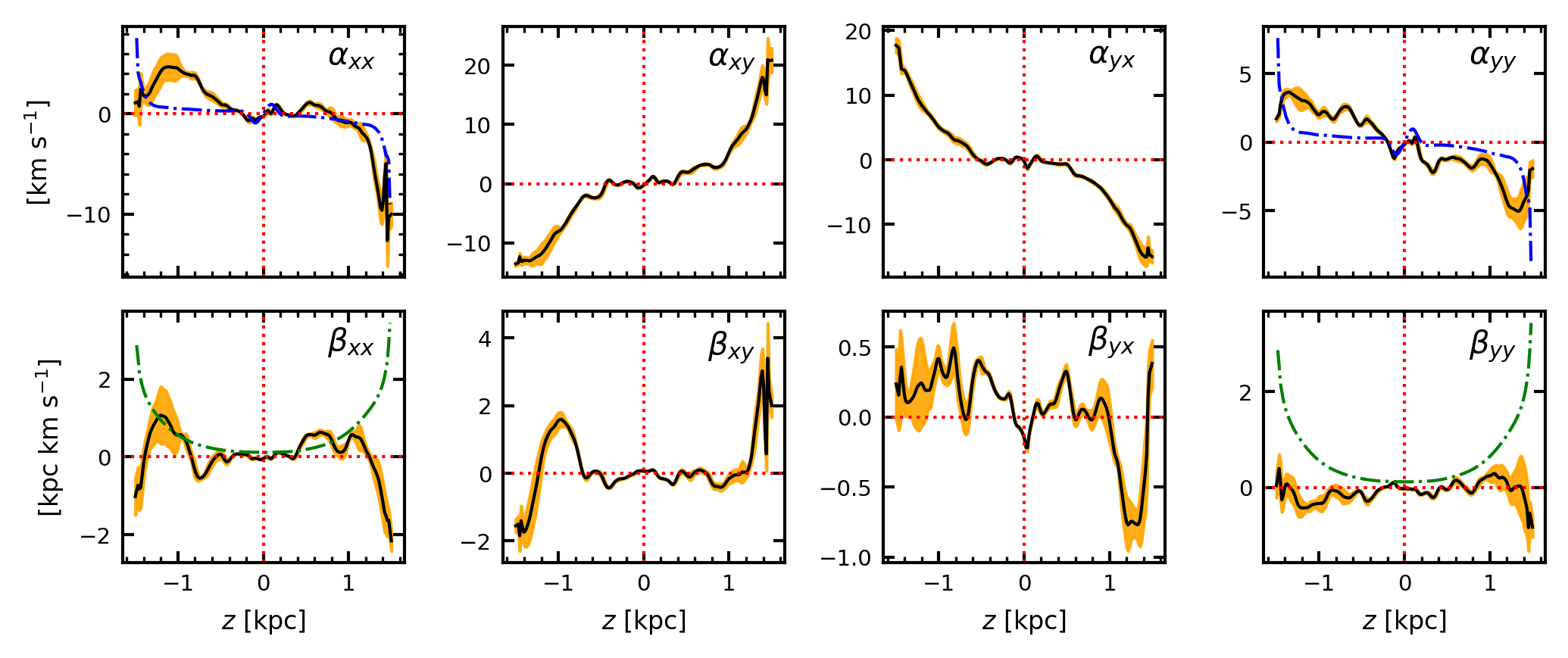}
    \caption{The elements of the turbulent transport tensors introduced in
equation~\eqref{EMF} for Model~\RXhA\ in the nonlinear state at $1.5\leq t\leq
3.0\Gyr$.  The yellow stripes indicate one standard deviation of the variables
based on bootstrap resampling  of the time series of $\vec{\mathcal{E}}$. The
blue dash-dotted line in the upper row of panels is the sum of the kinetic and
magnetic helicities $\alpha\kin(z)+\alpha\magn(z)$, obtained using
equations~\eqref{alphak} and \eqref{alpham}, and thus not including the imposed
$\alpha$ of equation~\eqref{eq:alpha}.  The green dash-dotted line in the lower
row of panels represents the total magnetic diffusivity $\eta_\text{t}+\eta$
obtained using equation~\eqref{tdiff}.
}
    \label{fig:SVD_nonlinear}
\end{figure*}

\subsection{Electromotive force and transport coefficients}\label{TEMF}

To verify, refine and further justify our interpretation of the results,
particularly the dynamo action driven by the MBI, we have computed the
components of the (pseudo-)tensor $\alpha_{ij}$ and tensor $\beta_{ij}$ using
the method of singular value decomposition (SVD) introduced by
\cite{ABendre2019} and \citet{ABendre2022}. In this approach, the time series
for the Cartesian components of the electromotive force
$\mathcal{E}_i=\mean{\vec{u}\times\vec{b}}_i$ are approximated by
$\mathcal{E}_i=\alpha_{ij}\mean{B_j}-\beta_{ij}(\nabla\times\mean{\vec{B}})_j$.
Explicitly,
\begin{equation}\label{EMF}
\begin{pmatrix} \mathcal{E}_x\\ \mathcal{E}_y\end{pmatrix}
=\begin{pmatrix} \alpha_{xx} &\alpha_{xy}\\ \alpha_{yx} &\alpha_{yy} \end{pmatrix}
\begin{pmatrix} \mean{B}_x \\ \mean{B}_y \end{pmatrix}
-\begin{pmatrix} \beta_{xx} &\beta_{xy}\\ \beta_{yx} &\beta_{yy} \end{pmatrix}
\begin{pmatrix} (\nabla\times\mean{\vec{B}})_x \\ (\nabla\times\mean{\vec{B}})_y \end{pmatrix}
\end{equation}
are solved to determine the elements of the tensors $\alpha_{ij}$ and
$\beta_{ij}$, which are assumed to be independent of time. This assumption is
valid in either the early stages of the exponential growth of the magnetic
field or in the later, stationary state of the system. In these calculations
horizontal averaging is used, $\mean{\vec{B}}=\meanh{\vec{B}}$, as displayed in
Fig.~\ref{fig:xy_Bx_By_ha_02_ra_10}, such that the tensor elements are
functions of $z$ alone.  The horizontal average of the vertical component of
the magnetic field  vanishes due to the horizontal periodic boundary
conditions.  Hence, the analysis is applied only to the horizontal components
of the magnetic field.

We also estimated the components of $\alpha_{ij}$ and $\beta_{ij}$ using the
alternative IROS method described by \citet{bendre2023iterative}. The results
obtained with the two independent methods are quite consistent with each other,
and we present the SVD results below.

The governing equation for the mean magnetic field (when the mean gas velocity
vanishes, as in our case) has the form
\begin{equation}\label{MFDeq}
\deriv{\mean{\vec{B}}}{t}=\nabla\times(\vec{\mathcal{E}}-\eta\nabla\times\mean{\vec{B}})\,.
\end{equation}
The diagonal elements of the $\alpha$-tensor represent the $\alpha$-effect,
with $\alpha\kin+\alpha\magn\approx\tfrac12(\alpha_{xx}+\alpha_{yy})$. If the
flow is isotropic in the $(x,y)$-plane $\alpha_{ij}$ is antisymmetric
($\alpha_{yx}=-\alpha_{xy}$), and the off-diagonal elements of $\alpha_{ij}$
represent the transfer of the mean magnetic field along the $z$-axis at the
effective speed $U_z=-\alpha_{xy}$ due to the increase in the turbulent
magnetic diffusivity with $|z|$ resulting mainly from the increase of the
random flow speed \citep[turbulent diamagnetism -- e.g., Section~7.9
of][]{SS21}.
The diagonal components of the tensor $\beta_{ij}$ represent the  turbulent magnetic diffusion. 
Since the imposed $\alpha$-effect of equation~\eqref{eq:alpha}
and the magnetic diffusivity $\eta$ in equation~\eqref{eq:induction} are
not associated with any explicit velocity and magnetic field components, 
the elements of $\alpha_{ij}$ 
and $\beta_{ij}$ obtained with the SVD method are not
sensitive to them.

Figure~\ref{fig:SVD_nonlinear} presents the resulting components of the tensors
$\alpha_{ij}$ and $\beta_{ij}$ for the nonlinear stage of the evolution.  The
yellow stripes represent one standard deviation of the variables obtained from
five estimates, each resulting from the sampling of every fifth entry in the
time series of $\vec{\mathcal{E}}$ containing 2500 data points at each $z$,
measured with the time interval $1\Myr$.

Confirming the results and arguments of Section~\ref{HFDBMF},
$\alpha_{xx}+\alpha_{yy}$ is significant in magnitude, antisymmetric with
respect to the midplane $z=0$ and mostly negative at $z>0$. The magnitudes of
$\alpha_{xx}$ and $\alpha_{yy}$ are close to $\alpha\kin+\alpha\magn$ obtained
using equations~\eqref{alphak} and \eqref{alpham} and increase with $|z|$.

The off-diagonal components of $\alpha_{ij}$ are quite close to the expected
antisymmetry, $\alpha_{yx}=-\alpha_{xy}$ and reflect the transfer of the mean
magnetic field towards $z=0$ associated with the increase of the turbulent
magnetic diffusivity with $|z|$, which opposes the buoyant migration of the
magnetic field away from the midplane. The estimated values of $\beta_{xx}$ and
$\beta_{yy}$ shown in the lower row of Fig.~\ref{fig:SVD_nonlinear} are
significantly larger than $\eta=0.03\kpc\kms$, which shows that the velocity
field develops a rather strong random component by $t=1.5\Gyr$.  The components
of $\beta_{ij}$ obtained with the SVD agree less with equation~\eqref{tdiff}
although $\beta_{xx}$ is reasonably consistent with the analytical estimate at
$|z|\lesssim1\kpc$. The accuracy of the SVD results is expected to be lower for
the magnetic diffusion tensor because they involve the noisier spatial
derivatives of the mean magnetic field.

To confirm our conclusion that the dynamo action and the associated complex
behaviour of the mean magnetic field are essentially nonlinear phenomena, we
have verified that the components of both $\alpha_{ij}$ and $\beta_{ij}$
fluctuate around the zero level during the linear stage of the evolution
without any significant effect on the evolution.

\section{Summary and implications}\label{sec:conc}
The nonlinear interaction of the mean-field dynamo and magnetic buoyancy leads
to profound changes in the evolution of the large-scale magnetic field
transforming its properties far beyond what can be envisaged from a study of
the early stages of the magnetic field amplification (when the Lorentz force is
still negligible). A buoyant magnetic field not only spreads naturally to
larger altitudes from a relatively thin layer where it is generated by the
dynamo, but the buoyancy can change dramatically the magnetic field that has
caused it, reversing its direction and leading to nonlinear magnetic
oscillations. Neither the dynamo system in a plane layer nor the magnetic
buoyancy instability exhibit any oscillatory behaviours on their own.

The nonlinear development of the magnetic buoyancy instability of an imposed
uni-directional magnetic field has been studied by \citet{DT2022a,DT2022b}. We
have extended those results to the case of a buoyant magnetic field supported
by the mean-field dynamo action. Both systems behave similarly as they reverse
the direction of the magnetic field, and we attribute this to the secondary
mean-field dynamo action driven by the helical flows associated with the
magnetic buoyancy.

The ability of magnetic buoyancy to produce the $\alpha$-effect and thus a
mean-field dynamo action is well known but its synergy with the dynamo action
of a turbulent flow driven independently has not been explored earlier.

The following picture emerges from the studies of \citet{DT2022a,DT2022b} and
our work.

When the buoyant magnetic field is not helical (e.g., uni-directional) and
there is no rotation, magnetic buoyancy just redistributes the large-scale
magnetic field to larger altitudes reducing very strongly its pressure gradient
and leaving the support of the gas layer against the gravity to the thermal
pressure gradient and contributions from turbulence and random magnetic fields
if present \citep{DT2022a}. Rotation changes the picture because the gas flows
that accompany magnetic buoyancy become helical driving a mean-field dynamo
that overwhelms the imposed magnetic field leading to its reversal, apparently
signifying the onset of nonlinear oscillations \citep{DT2022b}. As we show
here, similar oscillations occur independently of the rotation if only the
buoyant magnetic field is produced by the mean-field dynamo. In this case, the
magnetic field is helical, so the associated Lorentz force powers helical flows
independently of the rotation and its Coriolis force. Because of the
nonlinearity, flows over a wide range of scales develop and are randomized
\citep{Rodrigues2016}.

Notably, the sign of the $\alpha$-effect caused by the Lorentz force is
opposite to that of the conventional $\alpha$-effect resulting from the action
of the Coriolis force.

Our simulations are performed in a relatively large but still limited part of a gas layer ($2\times2\kpc$ horizontally) using Cartesian coordinates. The computational domain is large enough to accommodate the most rapidly growing MBI mode, and it is likely that the results would not be much different in cylindrical coordinates where the unstable magnetic field would no longer be unidirectional. Therefore, it is reasonable to expect that our main conclusions apply to galactic and accretion discs as a whole, at least at some distance from the disc axis where the curvature is not very strong.

These results can change our understanding of the nonlinear evolution of
large-scale magnetic fields in spiral galaxies and accretion discs. In the case
of the mean-field dynamos in disc galaxies, the conventional picture
\citep{SS21} is that of a non-oscillatory, monotonically growing large-scale
magnetic field whose direction is controlled by the initial conditions. The
strength of the magnetic field can decrease with time if the interstellar gas
is depleted \citep{Luiz_R_2015a} but this does not affect its direction.

Magnetic buoyancy and its transformation of the dynamo action enrich this
picture with the possibility that the magnetic field in a spiral galaxy or
accretion disc can become oscillatory at a later stage when its energy density
becomes comparable to the turbulent and thermal energy densities. Both the
strength and direction of the large-scale magnetic field can vary in time in
this state.  Such an oscillatory behaviour has been reported in the simulation
of the mean-field dynamo action by supernova-driven turbulence under
large-scale shear by \citet{GMK23}. In their case, a single sign reversal of
the mean magnetic field occurs close to dynamo saturation. This corresponds to
a change in the sign of $\alpha$, which is due primarily to a late increase in
the strength of $\alpha\magn$, so could be a signature of MBI in their system.

The intensity and consequences of the interaction of the magnetic dynamo and
buoyancy depend on the relative intensities of the two processes and may vary
with location within a galaxy and between galaxies. The dynamo efficiency is
controlled by the scale height of the gas and rotation frequency. The
consequences of magnetic buoyancy also depend on these parameters but in a
different manner. The implications of our findings for the large-scale magnetic
fields in astrophysical discs require a dedicated exploration.

\section*{Acknowledgements}

We would like to thank Graeme Sarson for useful discussions. We are grateful to the anonymous referee for useful comments which helped to improve the text. The authors
benefited from valuable discussions at the Nordita workshop ``Towards a
Comprehensive Model of the Galactic Magnetic Field'' at Nordita (Stockholm) in
2023, supported by NordForsk and Royal Astronomical Society. F.A.G.\
acknowledges support of the Ministry of Education and Culture Global Programme
USA Pilot 9758121, the ERC under the EU's Horizon 2020 research and innovation
programme (Project UniSDyn, grant 818665) and the Swedish Research Council
(Vetenskapsr\aa det) grant 2022-03767.

\section*{Data Availability}

The raw data for this work were obtained from numerical simulations using the
open source PENCIL-CODE available at
\url{https://github.com/pencil-code/pencil-code.git}. The derived data used for
the analysis given in the paper is available on request from the corresponding
author.


\bibliographystyle{mnras}
\bibliography{refs}

\begin{thebibliography}{}
\makeatletter
\relax
\def\mn@urlcharsother{\let\do\@makeother \do\$\do\&\do\#\do\^\do\_\do\%\do\~}
\def\mn@doi{\begingroup\mn@urlcharsother \@ifnextchar [ {\mn@doi@} {\mn@doi@[]}}
\def\mn@doi@[#1]#2{\def\@tempa{#1}\ifx\@tempa\@empty \href {http://dx.doi.org/#2} {doi:#2}\else \href {http://dx.doi.org/#2} {#1}\fi \endgroup}
\def\mn@eprint#1#2{\mn@eprint@#1:#2::\@nil}
\def\mn@eprint@arXiv#1{\href {http://arxiv.org/abs/#1} {{\tt arXiv:#1}}}
\def\mn@eprint@dblp#1{\href {http://dblp.uni-trier.de/rec/bibtex/#1.xml} {dblp:#1}}
\def\mn@eprint@#1:#2:#3:#4\@nil{\def\@tempa {#1}\def\@tempb {#2}\def\@tempc {#3}\ifx \@tempc \@empty \let \@tempc \@tempb \let \@tempb \@tempa \fi \ifx \@tempb \@empty \def\@tempb {arXiv}\fi \@ifundefined {mn@eprint@\@tempb}{\@tempb:\@tempc}{\expandafter \expandafter \csname mn@eprint@\@tempb\endcsname \expandafter{\@tempc}}}

\bibitem[\protect\citeauthoryear{{Baryshnikova} \& {Shukurov}}{{Baryshnikova} \& {Shukurov}}{1987}]{BaSh87}
{Baryshnikova} I.,  {Shukurov} A.,  1987, \mn@doi [Astron.\ Nachr.] {10.1002/asna.2113080202}, \href {https://ui.adsabs.harvard.edu/abs/1987AN....308...89B} {308, 89}

\bibitem[\protect\citeauthoryear{{Bendre} \& {Subramanian}}{{Bendre} \& {Subramanian}}{2022}]{ABendre2022}
{Bendre} A.~B.,  {Subramanian} K.,  2022, \mn@doi [\mnras] {10.1093/mnras/stac339}, \href {https://ui.adsabs.harvard.edu/abs/2022MNRAS.511.4454B} {511, 4454}

\bibitem[\protect\citeauthoryear{{Bendre}, {Subramanian}, {Elstner}  \& {Gressel}}{{Bendre} et~al.}{2020}]{ABendre2019}
{Bendre} A.~B.,  {Subramanian} K.,  {Elstner} D.,   {Gressel} O.,  2020, \mn@doi [\mnras] {10.1093/mnras/stz3267}, \href {https://ui.adsabs.harvard.edu/abs/2020MNRAS.491.3870B} {491, 3870}

\bibitem[\protect\citeauthoryear{Bendre, Schober, Dhang  \& Subramanian}{Bendre et~al.}{2023}]{bendre2023iterative}
Bendre A.~B.,  Schober J.,  Dhang P.,   Subramanian K.,  2023, Iterative removal of sources to model the turbulent electromotive force (\mn@eprint {arXiv} {2308.00059})

\bibitem[\protect\citeauthoryear{{Brandenburg} \& {Dobler}}{{Brandenburg} \& {Dobler}}{2002}]{brandenburg2002}
{Brandenburg} A.,  {Dobler} W.,  2002, \mn@doi [Comput.\ Phys.\ Commun.] {10.1016/S0010-4655(02)00334-X}, \href {https://ui.adsabs.harvard.edu/abs/2002CoPhC.147..471B} {147, 471}

\bibitem[\protect\citeauthoryear{{Brandenburg} \& {Sarson}}{{Brandenburg} \& {Sarson}}{2002}]{ABGS02}
{Brandenburg} A.,  {Sarson} G.~R.,  2002, \mn@doi [\prl] {10.1103/PhysRevLett.88.055003}, \href {https://ui.adsabs.harvard.edu/abs/2002PhRvL..88e5003B} {88, 055003}

\bibitem[\protect\citeauthoryear{{Brandenburg} \& {Subramanian}}{{Brandenburg} \& {Subramanian}}{2005}]{Brandenburg_2005}
{Brandenburg} A.,  {Subramanian} K.,  2005, \mn@doi [\physrep] {10.1016/j.physrep.2005.06.005}, \href {https://ui.adsabs.harvard.edu/abs/2005PhR...417....1B} {417, 1}

\bibitem[\protect\citeauthoryear{{Ferri{\`e}re}}{{Ferri{\`e}re}}{1998}]{Ferriere_1998}
{Ferri{\`e}re} K.,  1998, \mn@doi [\apj] {10.1086/305469}, \href {https://ui.adsabs.harvard.edu/abs/1998ApJ...497..759F} {497, 759}

\bibitem[\protect\citeauthoryear{{Gaburov}, {Johansen}  \& {Levin}}{{Gaburov} et~al.}{2012}]{GJL12}
{Gaburov} E.,  {Johansen} A.,   {Levin} Y.,  2012, \mn@doi [\apj] {10.1088/0004-637X/758/2/103}, \href {https://ui.adsabs.harvard.edu/abs/2012ApJ...758..103G} {758, 103}

\bibitem[\protect\citeauthoryear{{Gent}, {Shukurov}, {Fletcher}, {Sarson}  \& {Mantere}}{{Gent} et~al.}{2013}]{Gent_SN_ISM_1}
{Gent} F.~A.,  {Shukurov} A.,  {Fletcher} A.,  {Sarson} G.~R.,   {Mantere} M.~J.,  2013, \mn@doi [\mnras] {10.1093/mnras/stt560}, \href {https://ui.adsabs.harvard.edu/abs/2013MNRAS.432.1396G} {432, 1396}

\bibitem[\protect\citeauthoryear{{Gent}, {Mac Low}, {K{\"a}pyl{\"a}}, {Sarson}  \& {Hollins}}{{Gent} et~al.}{2020}]{GMKSH20}
{Gent} F.~A.,  {Mac Low} M.-M.,  {K{\"a}pyl{\"a}} M.~J.,  {Sarson} G.~R.,   {Hollins} J.~F.,  2020, \mn@doi [Geophys.\ Astrophys.\ Fluid Dyn.] {10.1080/03091929.2019.1634705}, \href {https://ui.adsabs.harvard.edu/abs/2020GApFD.114...77G} {114, 77}

\bibitem[\protect\citeauthoryear{{Gent}, {Mac Low}, {K{\"a}pyl{\"a}}  \& {Singh}}{{Gent} et~al.}{2021}]{GMKS21}
{Gent} F.~A.,  {Mac Low} M.-M.,  {K{\"a}pyl{\"a}} M.~J.,   {Singh} N.~K.,  2021, \mn@doi [\apjl] {10.3847/2041-8213/abed59}, \href {https://ui.adsabs.harvard.edu/abs/2021ApJ...910L..15G} {910, L15}

\bibitem[\protect\citeauthoryear{{Gent}, {Mac Low}  \& {Korpi-Lagg}}{{Gent} et~al.}{2023}]{GMK23}
{Gent} F.~A.,  {Mac Low} M.-M.,   {Korpi-Lagg} M.~J.,  2023, {Transition from small-scale to large-scale dynamo in a supernova-driven, multiphase medium} (\mn@eprint {arXiv} {2306.07051})

\bibitem[\protect\citeauthoryear{{Hanasz} \& {Lesch}}{{Hanasz} \& {Lesch}}{1998}]{1998Han&Les}
{Hanasz} M.,  {Lesch} H.,  1998, \aap, \href {https://ui.adsabs.harvard.edu/abs/1998A&A...332...77H} {332, 77}

\bibitem[\protect\citeauthoryear{{Hanasz}, {Otmianowska-Mazur}  \& {Lesch}}{{Hanasz} et~al.}{2002}]{HO-ML02}
{Hanasz} M.,  {Otmianowska-Mazur} K.,   {Lesch} H.,  2002, \mn@doi [\aap] {10.1051/0004-6361:20020228}, \href {https://ui.adsabs.harvard.edu/abs/2002A&A...386..347H} {386, 347}

\bibitem[\protect\citeauthoryear{{Hanasz}, {Kowal}, {Otmianowska-Mazur}  \& {Lesch}}{{Hanasz} et~al.}{2004}]{HKO-ML04}
{Hanasz} M.,  {Kowal} G.,  {Otmianowska-Mazur} K.,   {Lesch} H.,  2004, \mn@doi [\apjl] {10.1086/420697}, \href {https://ui.adsabs.harvard.edu/abs/2004ApJ...605L..33H} {605, L33}

\bibitem[\protect\citeauthoryear{{Johansen} \& {Levin}}{{Johansen} \& {Levin}}{2008}]{JoLe08}
{Johansen} A.,  {Levin} Y.,  2008, \mn@doi [\aap] {10.1051/0004-6361:200810385}, \href {https://ui.adsabs.harvard.edu/abs/2008A&A...490..501J} {490, 501}

\bibitem[\protect\citeauthoryear{{Kosi{\'n}ski} \& {Hanasz}}{{Kosi{\'n}ski} \& {Hanasz}}{2007}]{KH07}
{Kosi{\'n}ski} R.,  {Hanasz} M.,  2007, \mn@doi [\mnras] {10.1111/j.1365-2966.2007.11476.x}, \href {https://ui.adsabs.harvard.edu/abs/2007MNRAS.376..861K} {376, 861}

\bibitem[\protect\citeauthoryear{{Krause} \& {R{\"a}dler}}{{Krause} \& {R{\"a}dler}}{1980}]{1980mfmd_Krause_Radler}
{Krause} F.,  {R{\"a}dler} K.-H.,  1980, {Mean-Field Magnetohydrodynamics and Dynamo Theory}.
Pergamon Press (also Akademie-Verlag: Berlin), Oxford

\bibitem[\protect\citeauthoryear{{Kuijken} \& {Gilmore}}{{Kuijken} \& {Gilmore}}{1989}]{K&G1989MNRAS}
{Kuijken} K.,  {Gilmore} G.,  1989, \mn@doi [\mnras] {10.1093/mnras/239.2.571}, \href {https://ui.adsabs.harvard.edu/abs/1989MNRAS.239..571K} {239, 571}

\bibitem[\protect\citeauthoryear{Moffatt}{Moffatt}{1978}]{Moffatt1978}
Moffatt K.,  1978, The Generation of Magnetic Fields in Electrically Conducting Fluids.
Cambridge University Press, Cambridge

\bibitem[\protect\citeauthoryear{{Moss}, {Shukurov}  \& {Sokoloff}}{{Moss} et~al.}{1999}]{1999MOSS}
{Moss} D.,  {Shukurov} A.,   {Sokoloff} D.,  1999, \aap, \href {https://ui.adsabs.harvard.edu/abs/1999A&A...343..120M} {343, 120}

\bibitem[\protect\citeauthoryear{{Parker}}{{Parker}}{1966}]{1966ApJ_145_811P}
{Parker} E.~N.,  1966, \mn@doi [\apj] {10.1086/148828}, \href {https://ui.adsabs.harvard.edu/abs/1966ApJ...145..811P} {145, 811}

\bibitem[\protect\citeauthoryear{{Parker}}{{Parker}}{1979}]{Parker1979}
{Parker} E.~N.,  1979, {Cosmical Magnetic Fields: Their Origin and Their Activity}.
Clarendon Press, Oxford

\bibitem[\protect\citeauthoryear{{Parker}}{{Parker}}{1992}]{1992PARKER}
{Parker} E.~N.,  1992, \mn@doi [\apj] {10.1086/172046}, \href {https://ui.adsabs.harvard.edu/abs/1992ApJ...401..137P} {401, 137}

\bibitem[\protect\citeauthoryear{{Pencil Code Collaboration} et~al.,}{{Pencil Code Collaboration} et~al.}{2021}]{Pencil-JOSS}
{Pencil Code Collaboration} et~al., 2021, \mn@doi [J.\ Open Source Software] {10.21105/joss.02807}, \href {https://ui.adsabs.harvard.edu/abs/2021JOSS....6.2807P} {6, 2807}

\bibitem[\protect\citeauthoryear{{R{\"a}dler} \& {Br{\"a}uer}}{{R{\"a}dler} \& {Br{\"a}uer}}{1987}]{RaBr87}
{R{\"a}dler} K.-H.,  {Br{\"a}uer} H.-J.,  1987, \mn@doi [Astron.\ Nachr.] {10.1002/asna.2113080203}, \href {https://ui.adsabs.harvard.edu/abs/1987AN....308..101R} {308, 101}

\bibitem[\protect\citeauthoryear{{Rodrigues}, {Shukurov}, {Fletcher}  \& {Baugh}}{{Rodrigues} et~al.}{2015}]{Luiz_R_2015a}
{Rodrigues} L.~F.~S.,  {Shukurov} A.,  {Fletcher} A.,   {Baugh} C.~M.,  2015, \mn@doi [\mnras] {10.1093/mnras/stv816}, \href {https://ui.adsabs.harvard.edu/abs/2015MNRAS.450.3472R} {450, 3472}

\bibitem[\protect\citeauthoryear{{Rodrigues}, {Sarson}, {Shukurov}, {Bushby}  \& {Fletcher}}{{Rodrigues} et~al.}{2016}]{Rodrigues2016}
{Rodrigues} L.~F.~S.,  {Sarson} G.~R.,  {Shukurov} A.,  {Bushby} P.~J.,   {Fletcher} A.,  2016, \mn@doi [\apj] {10.3847/0004-637X/816/1/2}, \href {https://ui.adsabs.harvard.edu/abs/2016ApJ...816....2R} {816, 2}

\bibitem[\protect\citeauthoryear{{Shukurov} \& {Subramanian}}{{Shukurov} \& {Subramanian}}{2021}]{SS21}
{Shukurov} A.,  {Subramanian} K.,  2021, Astrophysical Magnetic Fields: From Galaxies to the Early Universe.
Cambridge University Press, Cambridge, \mn@doi{10.1017/9781139046657}

\bibitem[\protect\citeauthoryear{{Shukurov}, {Sokoloff}  \& {Ruzmaikin}}{{Shukurov} et~al.}{1985}]{ShSoRu85}
{Shukurov} A.~M.,  {Sokoloff} D.~D.,   {Ruzmaikin} A.~A.,  1985, Magnetohydrodynamics, \href {https://ui.adsabs.harvard.edu/abs/1985MagGi........9S} {21, 6}

\bibitem[\protect\citeauthoryear{{Sokoloff}, {Shukurov}  \& {Ruzmaikin}}{{Sokoloff} et~al.}{1983}]{Anvar1983}
{Sokoloff} D.,  {Shukurov} A.,   {Ruzmaikin} A.,  1983, \mn@doi [\gafd] {10.1080/03091928308221753}, \href {https://ui.adsabs.harvard.edu/abs/1983GApFD..25..293S} {25, 293}

\bibitem[\protect\citeauthoryear{{Tharakkal}, {Shukurov}, {Gent}, {Sarson}  \& {Snodin}}{{Tharakkal} et~al.}{2023a}]{DT2022b}
{Tharakkal} D.,  {Shukurov} A.,  {Gent} F.~A.,  {Sarson} G.~R.,   {Snodin} A.,  2023a, \mn@doi [\mnras] {10.1093/mnras/stad2475}, \href {https://ui.adsabs.harvard.edu/abs/2023MNRAS.525.2972T} {525, 2972}

\bibitem[\protect\citeauthoryear{{Tharakkal}, {Shukurov}, {Gent}, {Sarson}, {Snodin}  \& {Rodrigues}}{{Tharakkal} et~al.}{2023b}]{DT2022a}
{Tharakkal} D.,  {Shukurov} A.,  {Gent} F.~A.,  {Sarson} G.~R.,  {Snodin} A.~P.,   {Rodrigues} L. F.~S.,  2023b, \mn@doi [\mnras] {10.1093/mnras/stad2610}, \href {https://ui.adsabs.harvard.edu/abs/2023MNRAS.525.5597T} {525, 5597}

\bibitem[\protect\citeauthoryear{{Thelen}}{{Thelen}}{2000}]{Thelen_aw_2000}
{Thelen} J.~C.,  2000, \mn@doi [\mnras] {10.1046/j.1365-8711.2000.03420.x}, \href {https://ui.adsabs.harvard.edu/abs/2000MNRAS.315..165T} {315, 165}

\makeatother
\end{thebibliography}

\appendix

\section{Models with small magnetic Prandtl number}\label{App_Prm}
\begin{table}
\centering
\caption{As Table~\ref{tab:growth_rate_dynamo} but for $\nu=0.1\kpc\kms$ and
$\eta=0.03\kpc\kms$.
 }
\label{tab:growth_rate_dynamo_2}
\begin{adjustbox}{max width=0.45\textwidth}
\begin{tabular}{lccccccc}
\hline
Model  &R$_\alpha$ &$h_{\alpha}$ &$\alpha_0$   &$\gamma_\text{D}$ & $\gamma_\text{B}$ & $\gamma_u$  &  T  \\
       &           & [kpc]       &[km s$^{-1}$]&[Gyr$^{-1}$]      &       [Gyr$^{-1}$]& [Gyr$^{-1}$]& [Gyr]\\
\hline
\RThAb  &          3 & 0.2        & 0.12          &      -- & --     & --      & --  \\
\RThBb  &            & 0.3        & 0.08          &      -- & --     & --      & --  \\
\RThCb  &            & 0.6        & 0.04          &      -- &  --    & --      & --  \\[0.2cm]
\RVhAb  &          5 & 0.2        & 0.2           &    --   &  --    & --      & --  \\
\RVhBb  &            & 0.3        & 0.13          &    --   & --     & --      & --  \\
\RVhCb  &            & 0.6        & 0.06          &    --   & --     & --      & --  \\[0.2cm]
\RHhAb  &          7 & 0.2        & 0.28          &  1.01   &  2.9   & 2.1     & 1.5 \\
\RHhBb  &            & 0.3        & 0.186         &  0.49   &   --   & 1.0     & 1.9 \\
\RHhCb  &            & 0.6        & 0.093         &   --    &  --    & --      & --  \\[0.2cm]
\RXhAb  &         10 & 0.2        & 0.4           &  3.16   &   8.2  & 4.9     & 1.4 \\
\RXhBb  &            & 0.3        & 0.26          &  1.45   &   3.9  & 2.8     & 1.8 \\
\RXhCb  &            & 0.6        & 0.13          &  0.34   &   --   & 0.7     & --  \\[0.2cm]
\RXVhAb &         15 & 0.2        & 0.6           &  8.27   &   21.2 & 15.5    & 1.3 \\
\RXVhBb &            & 0.3        & 0.4           &  3.81   &   9.3  & 5.8     & 1.7 \\
\RXVhCb &            & 0.6        & 0.2           &  0.97   &   1.8  & 1.9     & 3.1 \\[0.2cm]
\RXXhAb &         20 & 0.2        & 0.8           & 15.26   & 42.3   & 29.8    & 1.4 \\
\RXXhBb &            & 0.3        & 0.53          & 7.0     & 16.7   & 10.0    & 1.9 \\
\RXXhCb &            & 0.6        & 0.26          & 1.8     &  6.2   & 3.4     & 3.0 \\[0.2cm] \hline
\end{tabular}
\end{adjustbox}
\end{table}

The simulations used in the main text have $\nu>\eta$, i.e.,
$\textrm{Pr}\magn=\nu/\eta>1$ in terms of the magnetic Prandtl number. Since
the MBI is sensitive to both kinematic viscosity $\nu$ and magnetic diffusivity
$\eta$, we also considered models with $\textrm{Pr}\magn=\nu/\eta<1$ with
parameters specified in Table~\ref{tab:growth_rate_dynamo_2}. Although we do
not discuss all the models presented in Tables~\ref{tab:growth_rate_dynamo} and
\ref{tab:growth_rate_dynamo_2} in the text, our conclusions are based on the
results obtained in all these models.

\section{Correlation time and length}
\label{appendix:cor_time_len}
To determine the correlation scale of a random variable $f(\vec{x})$, such as
the random speed $|\vec{u}|$, we first compute the second-order structure
function
\begin{equation}
    D(l) = \langle \lvert f(\vec{x} - \vec{l}) - f(\vec{x})\rvert^2 \rangle_{xy}\,.
    \label{eq:struct_func}
\end{equation}
where $l=|\vec{l}|$.  The systems considered here are statistically
homogeneous in the horizontal planes and stratified along the $z$-axis.
Therefore, we use horizontal averages, $\mean{\cdots}=\meanh{\cdots}$, and
consider lags confined to the $(x,y)$-plane, $\vec{l}=(l_x,l_y,0)$. Because of
the periodic boundary conditions in $x$ and $y$, we consider the ranges
$0<l_x<L_x/2$ and $0<\ell_y<L_y/2$, where $L_{x}$ and $L_{y}$ are the
corresponding sizes of the computational domain, $L_x=L_y=4\kpc$.  The
correlation length is obtained from the autocorrelation function $C(l)$,
\begin{equation}\label{l0}
l_0 = \int_0^\infty C(l)\,\dd l\,,
\qquad
C(l) = 1 - D(l)/(2\sigma^{2})\,,
\end{equation}
where $2\sigma^2$ is the value of $D(l)$  at large $l$ where the values of
$f(\vec{x})$ are no longer correlated. The choice of $\sigma^2$ is not always
obvious because of the limited size of the domain, so 
$\sigma$ and $l_0$ are obtained from the analytical fit of the following form to the values of $D(l)$ derived from the data:
\begin{equation}
D(l) = 2\sigma^{2}\left[1- \exp\left(- l^2/(2L_0^2)\right)\right], \quad l_0 = \sqrt{\pi/2}\,L_0\,,
        \label{eq:D_cor_len}
\end{equation}

This procedure equally applies to the time correlations, where $f$ is a
function of time $t$ rather than position $\vec{x}$, leading to an estimate of
the correlation time $\tau$. However,  instead of using equation~\eqref{eq:D_cor_len} we use
\begin{equation}
D(l) = 2\sigma^{2}\left[1- \exp\left(-l/L_0\right)\right], \quad l_0 = L_0\,,
    \label{eq:D_cor_time}
\end{equation}
 This difference is ultimately related to the fact that the governing equations
contain the first-order time derivatives and second-order spatial derivatives,
so that the random velocity and magnetic fields are continuous but not
necessarily smooth functions of time but smooth functions of position.

\bsp	
\label{lastpage}
\end{document}